\documentclass[12pt]{article}

\catcode`\@=11
\def\numberbysection{\@addtoreset{equation}{section}
\def\theequation{\thesection.\arabic{equation}}}
\numberbysection
\input epsf
\newcommand{\abs}[1]{\left\vert#1\right\vert}
\newcommand{\beq}{\begin{equation}}
\newcommand{\beqa}{\begin{eqnarray}}
\newcommand{\bin}[2]{{#1\choose#2}}
\newcommand{\C}{{\bf C}}
\newcommand{\capt}[3]{\noindent{\bf Figure~#1: }#3{\vskip .2cm}}
\renewcommand{\d}{\mathrm{d}}
\newcommand{\dbli}[2]{_{\matrix{\scriptstyle#1\cr
\noalign{\vskip -6pt}\scriptstyle#2}}}
\renewcommand{\deg}{\mathrm{deg}}
\newcommand{\e}{\mathrm{e}}
\newcommand{\eeq}{\end{equation}}
\newcommand{\eeqa}{\end{eqnarray}}
\newcommand{\eps}{\varepsilon}
\newcommand{\Frac}{\mathrm{Frac}}
\newcommand{\frad}[2]{\displaystyle{\displaystyle#1\over\displaystyle#2}}
\renewcommand{\i}{\mathrm{i}}
\newcommand{\Int}{\mathrm{Int}}
\newcommand{\m}{-}
\renewcommand{\max}{_{\mathrm{max}}}

\newcommand{\mean}[1]{\left\langle#1\right\rangle}
\newcommand{\N}{{\cal N}}
\renewcommand{\o}{\omega}
\renewcommand{\O}{{\cal O}}
\newcommand{\oA}{{\overline{A}}}
\newcommand{\oF}{{\overline{F}}}
\newcommand{\p}{+}
\newcommand{\prob}{\mathrm{Prob}}
\newcommand{\V}{\overline{V}}
\newcommand{\w}{\widetilde}
\newcommand{\z}{\zeta}
\begin{document}
\centerline{\Large\bf Statistics of persistent events}
\vspace{.3cm}
\centerline{\Large\bf in the binomial random walk:}
\vspace{.3cm}
\centerline{\Large\bf Will the drunken sailor hit the sober man?}
\vspace{1cm}
\centerline{\large
by M.~Bauer$^{a,}$\footnote{bauer@spht.saclay.cea.fr},
C.~Godr\`eche$^{b,}$\footnote{godreche@spec.saclay.cea.fr},
and J.M.~Luck$^{a,}$\footnote{luck@spht.saclay.cea.fr}
}
\vspace{1cm}
\centerline{$^a$ Service de Physique Th\'eorique,}
\centerline{CEA Saclay, 91191 Gif-sur-Yvette cedex, France}
\vspace{.5cm}
\centerline{$^b$ Service de Physique de l'\'Etat Condens\'e,}
\centerline{CEA Saclay, 91191 Gif-sur-Yvette cedex, France}
\vspace{1cm}
\begin{abstract}
The statistics of persistent events,
recently introduced in the context of phase ordering dynamics,
is investigated in the case of the one-dimensional lattice random walk
in discrete time.
We determine the survival probability of the random walker
in the presence of an obstacle moving ballistically with velocity $v$,
i.e., the probability that the random walker remains
up to time $n$ on the left of the obstacle.
Three regimes are to be considered for the long-time behavior
of this probability,
according to the sign of the difference between $v$ and
the drift velocity $\V$ of the random walker.
In one of these regimes ($v>\V)$,
the survival probability has a non-trivial limit at long times,
which is discontinuous at all rational values of $v$.
An algebraic approach allows us to compute these discontinuities,
as well as several related quantities.
The mathematical structure underlying the solvability of this model
combines elementary number theory, algebraic functions,
and algebraic curves defined over the rationals.
\end{abstract}
\vfill
\noindent To be submitted for publication to Journal of Statistical Physics
\hfill T/98/124
\vskip -6pt
\noindent P.A.C.S.: 05.40.+j, 01.50.-r, 02.50.Ey.
\hfill S/98/086
\newpage
\setcounter{footnote}{0}
\section{Introduction}

Consider an asymmetric binomial random walk on a one-dimensional lattice.
In units of the lattice spacing,
the steps $\eps_m$ performed by the walker at integer times $t=m$
are independent identically distributed random variables, with the binary law
\beq
\eps_m=\left\{\matrix{
-1&\hbox{with probability}&p,\hfill\cr
+1&\hbox{with probability}&q=1-p.\hfill\cr
}\right.%}
\eeq

This paper is devoted to the analysis of the probability $F(n,v)$
that the walker remains, up to time $n$, on the left of an obstacle
moving ballistically with velocity $v$.
In a pictorial language, $F(n,v)$ is the probability that a sober man
walking at constant speed $v$,
and a drunken sailor stepping forward and backward
erratically, both leaving a pub at some initial time,
do not meet up to time $n$.
Alternatively, $F(n,v)$ can be viewed as
the survival probability of the walker in the presence of the obstacle.
Surprisingly, as we shall see, this probability is highly non-trivial,
especially as far as its $v$-dependence is concerned.

The quantity $F(n,v)$ also represents the probability that
the path of the random walker in the $(t,x)$-plane remains,
up to the integer time $t=n$, on the left of the straight wall $x=vt$.
Figures~1 to 3 provide illustrations of these definitions.

$$\centerline{\epsfbox{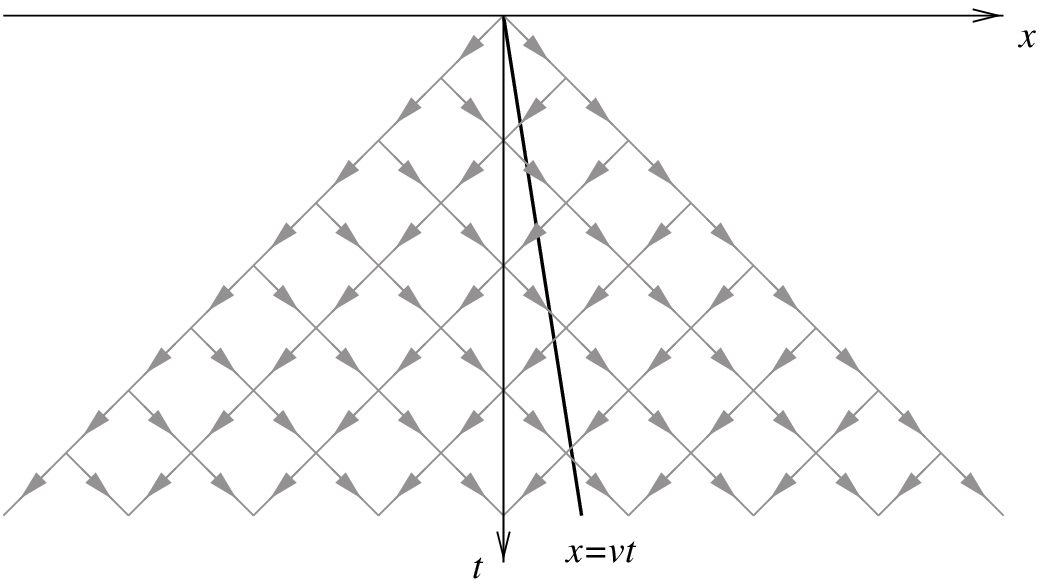}}$$
\capt{1}{M1}{Configuration space of the problem.}

This investigation was motivated by recent work on
the statistics of persistent events in
nonequilibrium statistical-mechanical systems
undergoing phase ordering~\cite{DG}.
Persistent events are defined by a constraint
on the past history of the system, to be satisfied up to time $t$,
i.e., for all previous times $0\le t'\le t$.
This concept can be simply illustrated on the example of a chain
of Ising spins, starting from a random initial condition, and evolving
under Glauber or heat-bath dynamics at zero temperature~\cite{Glauber}.
The persistence probability $R(t)$ is defined as the fraction
of space which remained in the same phase up to time $t$,
i.e., in the present case as the fraction of spins which did not flip
up to time~$t$~\cite{Der}.
For long times the persistence probability decays as
\beq
R(t)\sim t^{-\theta},
\eeq
where $\theta$ is the persistence exponent.

$$\centerline{\epsfbox{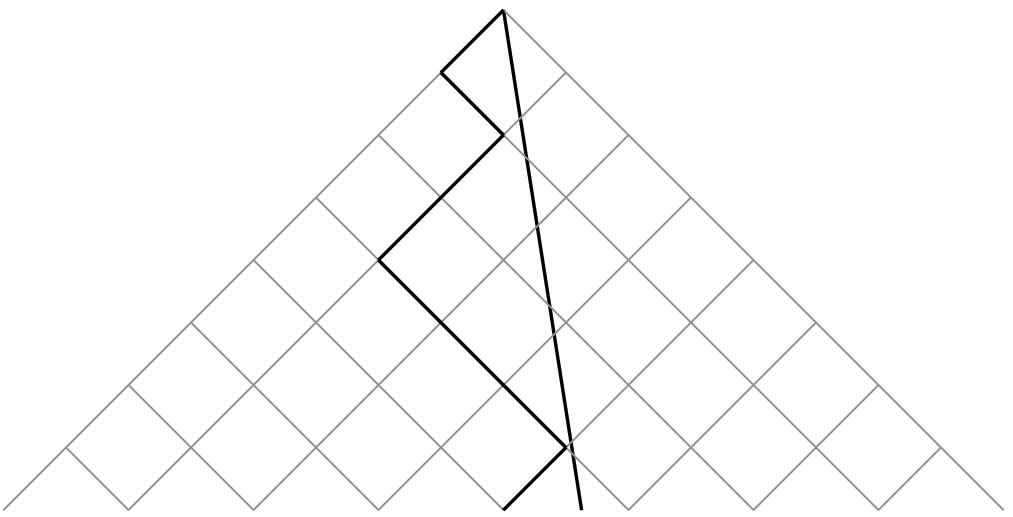}}$$
\capt{2}{M2}{A walk remaining on the left of the wall,
thus contributing to the survival probability.}

$$\centerline{\epsfbox{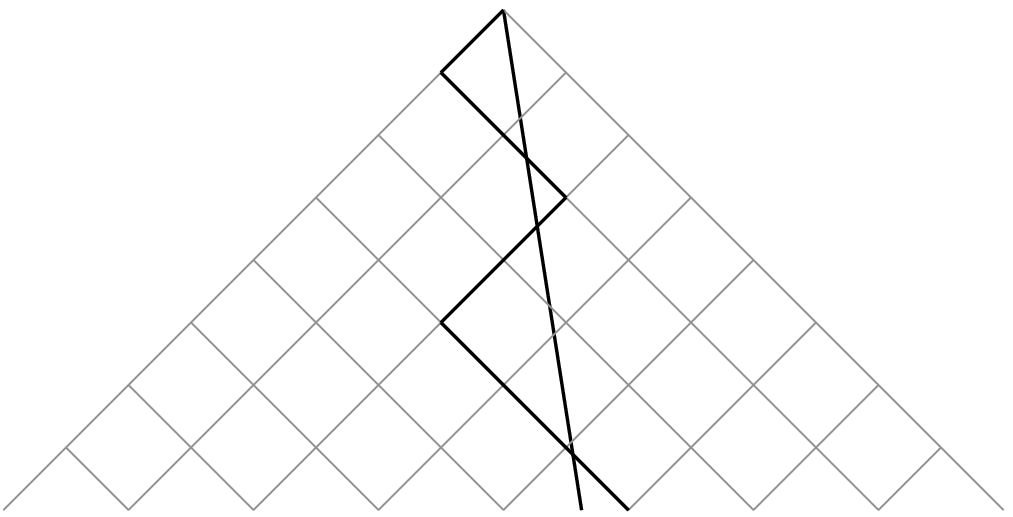}}$$
\capt{3}{M3}{A walk crossing the wall,
thus not contributing to the survival probability.}

The persistence exponent is actually only one member of a continuous
family of exponents defined as follows~\cite{DG}.
Denoting by $\sigma(t)$ the spin at a given site,
consider the local mean magnetization
\beq
M(t)=\frac{1}{t}\int_0^t\sigma(t')\d t'.
\eeq
This quantity is simply related to the fraction of time spent by
the spin in the positive direction.
Consider now the probability of {\it persistent large
deviations} $R(t,x)$ above the level $x$
(with $-1\le x\le 1)$, defined as the probability that
$M(t)$ remained greater than $x$, for all times $0\le t'\le t$.
For the Glauber-Ising chain at zero temperature, this quantity is
observed to decay algebraically at large times, with an exponent
$\theta(x)$ continuously varying with $x$~\cite{DG}.
When $x=1$, the usual persistence probability is recovered,
so that $\theta(1)=\theta$.
The exponent $\theta(x)$ appears as a first-passage exponent
associated to persistent large deviations of $M(t)$~\cite{DG,BBDG,Drouffe}.

The interpretation of the stochastic process $\sigma(t)$ as the
steps of a fictitious random walker naturally led us to
the present study, where, for simplicity, we consider
the concepts introduced above in the simplest case of a binomial random walk.
In the present work, the analogue of $M(t)$ is the mean velocity $V_n$
of the walker at time $n$,
\beq
V_n=\frac{x_n}{n},
\eeq
where
\beq
x_n=\sum_{m=1}^n\eps_m
\eeq
is the position of the walker at time $n$,
while $F(n,v)$ plays the role of the probability of persistent large deviations
$R(t,x)$.
Note a slight difference in the above definitions.
In this work, $F(n,v)$ is the probability that the mean speed was always
{\it less} than $v$, while in refs.~\cite{DG,BBDG,Drouffe}
$R(t,x)$ is the probability that the mean
magnetization was always {\it above} $x$.
This difference is harmless,
as it just amounts to interchanging the probabilities $p$ and $q$
and changing the slope $v$ into its opposite.

More precise definitions of the quantities considered in this work
are as follows.
First, the `one-time' distribution functions
$P^\pm(n,v)$ of the mean velocity $V_n$ {\it at} time $n$
are defined for a given slope $v$ as
\beqa
P^\m(n,v)&=&\prob\big\{V_n<v\big\}=\prob\big\{x_n<nv\big\},\nonumber\\
P^\p(n,v)&=&\prob\big\{V_n\le v\big\}=\prob\big\{x_n\le nv\big\},
\label{jpdef}
\eeqa
Then the survival probabilities
$F^\pm(n,v)$ {\it up to} time $n$ are defined as
\beqa
F^\m(n,v)&=&\prob\big\{V_m<v\,\hbox{ for }\,m=1,\dots,n\big\}
=\prob\big\{x_m<mv\,\hbox{ for }\,m=1,\dots,n\big\},\nonumber\\
F^\p(n,v)&=&\prob\big\{V_m\le v\,\hbox{ for }\,m=1,\dots,n\big\}
=\prob\big\{x_m\le mv\,\hbox{ for }\,m=1,\dots,n\big\}.\nonumber\\
\label{jfdef}
\eeqa
Two definitions are needed for each quantity,
since $V_n$ is a discrete random variable,
so that $P^\pm(n,v)$ and $F^\pm(n,v)$ are in general discontinuous
at any rational value of the slope $v$.
Throughout this paper, quantities with the $\p$ and the $\m$ superscript
are distinct from each other only when the slope $v$ is rational.
Equations involving $P^\pm(n,v)$, or $P(n,v)$ for short,
are meant to hold separately for $P^\p(n,v)$ and $P^\m(n,v)$.

We find three regimes for the
long-time behavior of $P^\pm(n,v)$ and $F^\pm(n,v)$,
according to the sign of the relative velocity of the obstacle
with respect to the drift velocity of the random walker
\beq
\V=\lim_{n\to\infty}V_n=\mean{\eps}=1-2p,
\label{jdrift}
\eeq
with the following results.

\begin{itemize}
\item{Large-deviation regime ($v<\V$):

Both $P^\pm(n,v)$ and $F^\pm(n,v)$ decay exponentially in time,
with a common entropy function $S(v)$,
but different power-law prefactors, namely
\beqa
&&P^\pm(n,v)\approx a_n^\pm(v)\,n^{-1/2}\,\e^{-nS(v)},\label{jregp1}\\
&&F^\pm(n,v)\approx b_n^\pm(v)\,n^{-3/2}\,\e^{-nS(v)}.\label{jregf1}
\eeqa
}
\item{Marginal regime ($v=\V$):

We have
\beqa
&&P^\pm(n,v)\to\frac{1}{2},\label{jregp2}\\
&&F^\pm(n,v)\approx\frac{C^\pm(v)}{(\pi n)^{1/2}}.\label{jregf2}
\eeqa
The limit value $1/2$ of $P^\pm(n,v)$
is a consequence of the central limit theorem.
The inverse-square-root decay of $F^\pm(n,v)$ is a known property
of one-dimensional random walks~\cite{F,DG,BBDG}.
}
\item{Convergent regime ($v>\V$):

$P^\pm(n,v)$ converge exponentially to one,
while $F^\pm(n,v)$ admit non-trivial limits:
\beqa
&&1-P^\pm(n,v)\approx a_n^\pm(v) n^{-1/2}\,\e^{-nS(v)},\label{jregp3}\\
&&{\hskip 7truemm}F^\pm(n,v)\to F^\pm(v).\label{jregf3}
\eeqa
}
\end{itemize}

These results were partially announced in ref.~\cite{DG}.
The asymptotic behavior of the one-time distribution functions $P^\pm(n,v)$
are well-known properties of random walks~\cite{F},
while the long-time behavior of the survival probabilities $F^\pm(n,v)$
in these three regimes is the subject of the present study.
To obtain this long-time behavior, we shall proceed in two steps.

A first level of description is provided in sec.~\ref{sec:sa}.
We use a combinatorial result, originally due to Sparre Andersen,
and proved in Appendix~A,
in order to derive the form of the estimates~(\ref{jregf1}), (\ref{jregf2}),
and~(\ref{jregf3}).
The prefactors $b_n^\pm(v)$ and $C^\pm(v)$,
as well as the limit survival probabilities $F^\pm(v)$,
are respectively given by eqs.~(\ref{jb}), (\ref{jc}), and~(\ref{jf}).
Such expressions hold for random walks with any distribution of steps $\eps_m$,
whose first two moments are finite.
However they are only formal, in the sense that they
do not lead to closed-form results in general.

A second, more refined level of description is provided by
the rest of the paper, devoted to a detailed study
of the quantities $b_n^\pm(v)$, $C^\pm(v)$, and $F^\pm(v)$.
This investigation will allow us to unravel an unexpected richness
in the statistics of persistent events in the present situation
of the binomial random walk on the lattice of integers.
More precisely, sec.~\ref{sec:prel} contains basic concepts and results,
emphasizing the central importance of sequences of positive integers $A^\pm_k$.
Sec.~\ref{sec:arb} is devoted to the exposition of several methods
(continuity of the path, probability flow, duality symmetry),
which allow the determination of the integers $A^\pm_k$ and related quantities,
in the case of an arbitrary slope $v$.
The case of a rational slope $v$ is dealt with in sec.~\ref{sec:algtrick},
where algebraic functions play a central role.
Sec.~\ref{sec:algcrit} contains an investigation of the critical behavior
of the survival probabilities,
obtaining thus predictions in the three regimes.
Further results on several specific situations,
including the slopes of the form $v=1-2/N$ or $v=-1+2/N$,
are derived in sec.~\ref{sec:further},
while sec.~\ref{sec:dis} contains a discussion.

\section{Sparre Andersen formalism}
\label{sec:sa}

As announced in the Introduction,
this section is devoted to a first level of description
of the one-time distribution functions $P^\pm(n,v)$ and
of the survival probabilities $F^\pm(n,v)$.

We shall make an extensive use of a remarkable combinatorial result,
originally due to Sparre Andersen~\cite{SA}.
The possibility of recasting the computation of $F^\pm(n,v)$
as a Sparre Andersen problem was uncovered in ref.~\cite{BBDG}.
The result of Sparre Andersen is best expressed as an identity
between the generating series of the $P^\pm(n,v)$ and of the $F^\pm(n,v)$,
namely
\beq
f^\pm(z,v)=\sum_{n\ge0}F^\pm(n,v)z^n
=\exp\left(\sum_{n\ge1}\frac{P^\pm(n,v)}{n}z^n\right).
\label{jic}
\eeq
This result can be found e.g. in the book by Feller~\cite{F}.
Since eq.~(\ref{jic}) lacks a clear intuitive meaning,
we give an elementary and self-contained combinatorial proof
of it in Appendix~A.

Taking the logarithmic derivative of both sides of eq.~(\ref{jic}),
we obtain linear recursion relations for $F^\pm(n,v)$:
\beq
nF^\pm(n,v)=\sum_{m=0}^{n-1}P^\pm(n-m,v)F^\pm(m,v),
\label{jpfrecur}
\eeq
with $F^\pm(0,v)=1$, hence
\beqa
F^\pm(1,v)&=&P^\pm(1,v),\nonumber\\
F^\pm(2,v)&=&\frac{1}{2}\left((P^\pm(1,v))^2+P^\pm(2,v)\right),\nonumber\\
F^\pm(3,v)&=&\frac{1}{6}
\left((P^\pm(1,v))^3+3P^\pm(1,v)P^\pm(2,v)+2P^\pm(3,v)\right),
\eeqa
and so on.

In the present situation,
the one-time distribution functions $P^\pm(n,v)$ can be evaluated
in closed form as follows.
We introduce the notations
\beq
p_c=\frac{1-v}{2},\qquad q_c=1-p_c=\frac{1+v}{2}.
\label{jpq}
\eeq
Let $k$ (respectively, $\w k$) be the number of steps to the left
(respectively, to the right),
i.e., the number of $\eps_m$ equal to $-1$ (respectively, to $+1$),
among the first $n$ steps.
The probability distribution of the integer $k=0,\dots,n$ reads
\beq
p_{n,k}=\bin{n}{k}p^kq^{n-k}.
\label{jpbin}
\eeq
The position $x_n$ of the particle at time $n$ is related to $k$ and $\w k$ by
\beq
n=k+\w k,\quad x_n=\w k-k=n-2k,
\quad k=\frac{n-x_n}{2},\quad\w k=\frac{n+x_n}{2}.
\label{jdefk}
\eeq
The condition $V_n<v$ (respectively, $V_n\le v$)
is equivalent to $k>np_c$ (respectively, $k\ge np_c$), hence the result
\beq
P^\pm(n,v)=\sum_{k=\Int^\pm(np_c)+1}^n p_{n,k},
\label{jpsum}
\eeq
where we have defined the following two
integer-part and fractional-part functions:
\beq
x=\Int^\pm(x)+\Frac^\pm(x),
\,\hbox{ with }\,\Int^\pm(x)\,\hbox{ integer and }
\,\left\{\matrix{0\le\Frac^\m(x)<1,\cr0<\Frac^\p(x)\le1.}\right.%}
\label{jdefif}
\eeq

The random walk possesses a duality symmetry,
defined by interchanging the probabilities $p$ and $q$,
and simultaneously changing the slope $v$ into its opposite $-v$,
i.e., interchanging $p_c$ and $q_c$.
This amounts to considering the survival probability
in the complementary domain,
i.e., on the right of the wall shown in Figure~1.
Let us make the dependence of quantities on $p$ explicit for a while,
indicating the value of $p$ after a semi-colon.
Our conventions imply that the one-time distribution functions
pertaining to the left of the wall
and to the right of the wall read respectively $P^\pm(n,v;p)$
and $1-P^\mp(n,-v;q)$.
These quantities are clearly equal:
\beq
P^\pm(n,v;p)+P^\mp(n,-v;q)=1.
\eeq
As a consequence,
the generating series $f^\pm(z,v)$ obeys the following identity:
\beq
f^\pm(z,v;p)\,f^\mp(z,-v;q)=\frac{1}{1-z}.
\label{jdusig}
\eeq
More spectacular consequences of this duality symmetry will be investigated in
secs.~\ref{sec:du} and~\ref{sec:algtrickdual},
by means of more powerful techniques.

We now turn to the investigation of the asymptotic behavior
of the survival probabilities.
As mentioned above, three regimes are to be considered,
according to the sign of the relative velocity of the obstacle
with respect to the drift velocity of the random walker
[see eq.~(\ref{jdrift})], namely
\beq
v-\V=v-(1-2p)=2(p-p_c).
\eeq

\subsection{Large-deviation regime: $v<\V$ or $p<p_c$}

This situation is typical of large deviations.
The chance for the mean velocity of the random walker to deviate
from its average is exponentially decreasing with time.
More precisely, a careful treatment of the sums in eq.~(\ref{jpsum}),
using eq.~(\ref{jpbin}) and Stirling's formula, leads to
\beq
P^\pm(n,v)\approx a_n^\pm(v)\,n^{-1/2}\,\e^{-nS(v)}\qquad(v<\V)
\label{jasyp}
\eeq
[see eq.~(\ref{jregp1})].
The associated entropy function (or large-deviation function) reads
\beqa
S(v)&=&p_c\ln\frac{p_c}{p}+q_c\ln\frac{q_c}{q}\nonumber\\
&=&\frac{1}{2}\left((1-v)\ln\frac{1-v}{2p}+(1+v)\ln\frac{1+v}{2q}\right).
\label{js}
\eeqa
This function is positive, and it vanishes quadratically as $p\to p_c$,
i.e., $\V\to v$, in agreement with the central limit theorem, according to
\beq
S(v)\approx\frac{(p-p_c)^2}{2p_cq_c}=\frac{(\V-v)^2}{2(1-v^2)}.
\eeq
In eq.~(\ref{jasyp}), the prefactors $a_n^\pm(v)=a^\pm(np_c)$
are $v$-dependent periodic functions of $np_c$, with unit period,
which can be determined explicitly:
\beq
a^\pm(x)=\frad{\left(\frad{q_cp}{p_cq}\right)^{1-\Frac^\pm(x)}}
{(2\pi p_cq_c)^{1/2}\left(1-\frad{q_cp}{p_cq}\right)},
\label{japer}
\eeq
with the definition~(\ref{jdefif}).
The amplitude functions $a^\pm(x)$ can be alternatively
expanded as Fourier series:
\beq
a^\pm(x)=\sum_{\ell=-\infty}^\infty\w a_\ell^\pm\e^{2\pi\i\ell x},
\eeq
so that eq.~(\ref{jasyp}) can be recast as
\beq
P^\pm(n,v)\approx n^{-1/2}\sum_{\ell=-\infty}^\infty
\w a_\ell^\pm\e^{\bigl(2\pi\i\ell p_c-S(v)\bigr)n}.
\label{jfoup}
\eeq

We remind for further reference the following general result,
known as a Tauberian theorem~\cite{F}.
If $c_n$ are positive numbers, with asymptotic behavior
\beq
c_n\approx a\,n^{\gamma}\,(z_c)^{-n}\qquad(n\gg1),
\label{tauc}
\eeq
then the power series
\beq
f(z)=\sum_{n\ge0}c_nz^n
\eeq
defines an analytic function whose nearest singularity is at $z=z_c$,
where it has a power-law singular part of the form
\beq
f_{\mathrm{sg}}(z)\approx a\,\Gamma(\gamma+1)
\,\left(1-\frac{z}{z_c}\right)^{-\gamma-1}\qquad(z\to z_c-0),
\label{tauf}
\eeq
$\Gamma$ denoting Euler's gamma function.
The reciprocal property, namely that eq.~(\ref{tauf}) implies eq.~(\ref{tauc}),
holds if the sequence $c_n$ is assumed to be smooth enough.

As a consequence of eq.~(\ref{tauf}), the estimate~(\ref{jfoup})
implies that the series $f^\pm(z,v)$ of eq.~(\ref{jic})
have square-root branch points at $z_\ell=\exp(S(v)-2\pi\i\ell p_c)$,
where $\ell$ runs over the integers.
Using eq.~(\ref{tauc}), we obtain the estimate
\beq
F^\pm(n,v)\approx b_n^\pm(v)\,n^{-3/2}\,\e^{-nS(v)}\qquad(v<\V),
\label{jasyf}
\eeq
with $b_n^\pm(v)=b^\pm(np_c)$,
where $b^\pm(x)$ are again periodic functions, with unit period,
whose Fourier coefficients are related to those of $a^\pm(x)$ by
\beq
\w b_\ell^\pm=\w a_\ell^\pm\,f^\pm(z_\ell).
\label{jb}
\eeq

The survival probabilities $F^\pm(n,v)$ thus decay exponentially
in the large-deviation regime,
with the same entropy function as the one-time distribution functions
$P^\pm(n,v)$, but with a different power of $n$
[see eqs.~(\ref{jregp1}),~(\ref{jregf1})],
multiplied by non-trivial periodic prefactors $b^\pm(np_c)$.

\vskip 8.5cm{\hskip 0.8cm}
\includegraphics{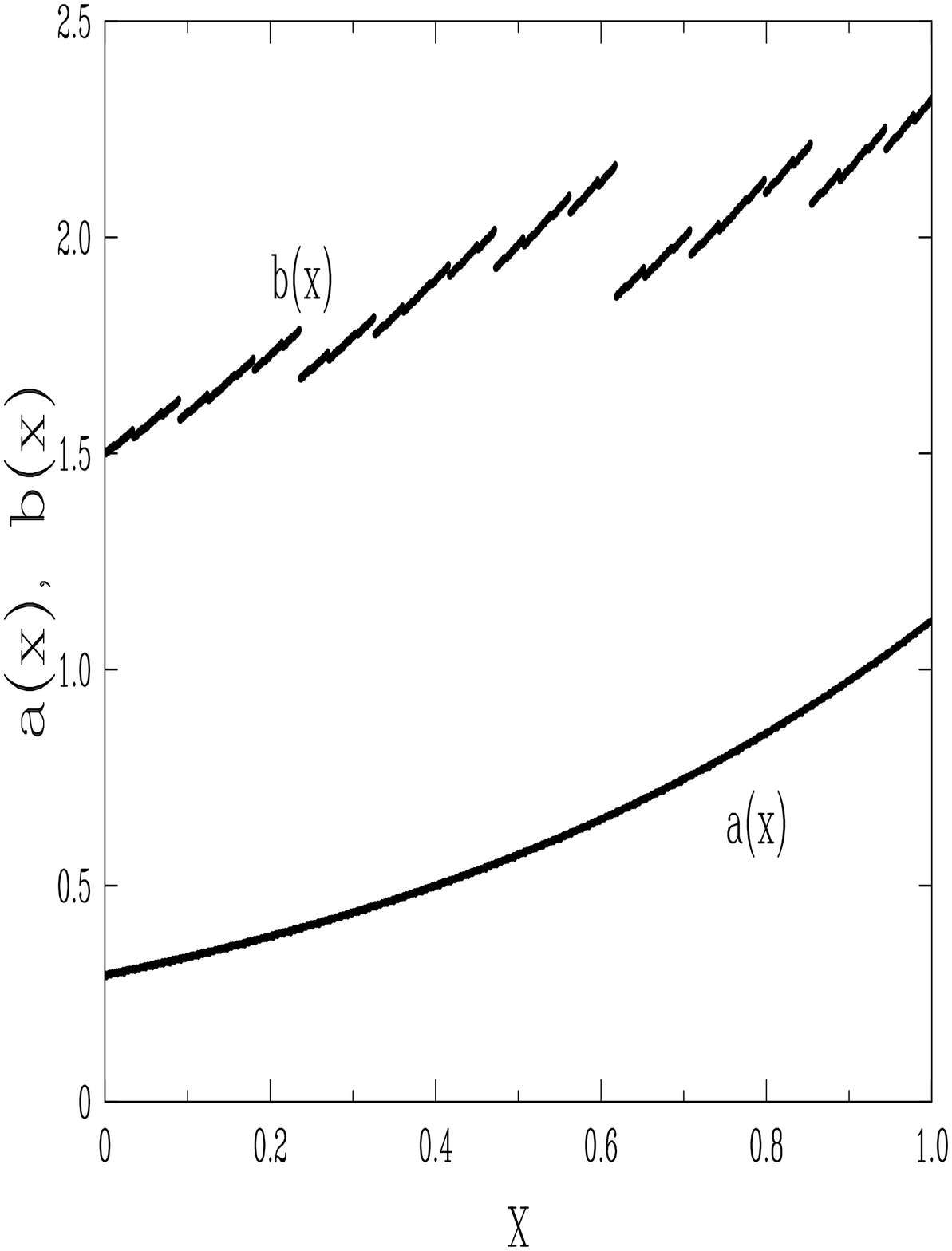}

\capt{4}{J4}{Periodic amplitudes $a(x)$ and $b(x)$,
for $p=0.3$ and $v$ equal to the golden slope.}

Figure~4 shows a plot of the periodic amplitudes $a(x)$ and $b(x)$,
for $p=0.3$, and for the `golden slope' $v$,
such that $p_c$ is the inverse golden mean, the most typical irrational:
\beq
p_c=\frac{\sqrt5-1}{2}=0.618034,\qquad v=2-\sqrt5=-0.236068.
\label{jgold}
\eeq
The data for $P(n,v)$ have been obtained by means of eq.~(\ref{jpsum}),
and for $F(n,v)$ by solving the recursion relation~(\ref{jpfrecur}).
The amplitude $a(x)$ is found to be a pure exponential,
in quantitative agreement with the prediction~(\ref{japer}).
The amplitude $b(x)$ exhibits a rich structure,
with discontinuities at all positive integer multiples of $p_c$ modulo 1.
The occurrence of these discontinuities can be explained as follows.
Eq.~(\ref{jpfrecur}) implies that
\beq
nF^\pm(n,v)=P^\pm(n,v)+F^\pm(1,v)P^\pm(n-1,v)+F^\pm(2,v)P^\pm(n-2,v)+\cdots
\label{nfasy}
\eeq
is a linear combination of $P^\pm(n,v)$, $P^\pm(n-1,v)$, $P^\pm(n-2,v)$,
and so on, with $n$-independent coefficients.
If we replace in eq.~(\ref{nfasy}) $P^\pm(n,v)$, $P^\pm(n-1,v)$,
$P^\pm(n-2,v)$, by their asymptotic form~(\ref{jasyp}),
with $a^\pm_n(v)=a^\pm(np_c)$,
we formally recover the result~(\ref{jasyf}), with
\beq
b^\pm(x)=a^\pm(x)+F^\pm(1,v)\,\e^{S(v)}\,a^\pm(x-p_c)
+F^\pm(2,v)\,\e^{2S(v)}\,a^\pm(x-2p_c)+\cdots
\label{bas}
\eeq
Since the amplitude functions $a^\pm(x)$, given by eq.~(\ref{japer}),
are discontinuous at $x=0$ modulo 1,
eq.~(\ref{bas}) shows that the amplitude functions $b^\pm(x)$ are
discontinuous at all positive integer multiples of $p_c$ modulo 1,
confirming thus the observation made on Figure~4.

\subsection{Marginal regime: $v=\V$ or $p=p_c$}
\label{sec:crit}

In this regime, the drift velocity $\V$ of the random walker
is equal to the velocity $v$ of the obstacle.
The one-time distribution functions $P^\pm(n,v)$ are expected to go to $1/2$,
as a consequence of the central limit theorem [see eq.~(\ref{jregp2})].

A careful treatment of eq.~(\ref{jpsum}) leads to the more accurate estimate
\beq
P^\pm(n,v)\approx\frac{1}{2}
+\left(\Frac^\pm(np_c)+\frac{p_c-2}{3}\right)\,(2\pi np_cq_c)^{-1/2}
\qquad(n\gg1).
\eeq
The Sparre Andersen relation~(\ref{jic}) then leads to the behavior~\cite{F}
\beq
f^\pm(z,v)\approx\frac{C^\pm(v)}{(1-z)^{1/2}}\qquad(z\to1),
\eeq
with
\beq
C^\pm(v)=\exp\left(\sum_{n\ge1}\frac{P^\pm(n,v)-1/2}{n}\right).
\label{jc}
\eeq
Eq.~(\ref{tauc}) in turn leads to
\beq
F^\pm(n,v)\approx\frac{C^\pm(v)}{(\pi n)^{1/2}}\qquad(v=\V).
\label{jfc}
\eeq

The survival probabilities $F^\pm(n,v)$ thus decay according to
a universal inverse-square-root law in the marginal case,
as already mentioned in eq.~(\ref{jregf2}).
The associated amplitudes $C^\pm(v)$ are non-trivial functions of $v$,
or equivalently of $p_c$,
to which we shall come back in sec.~\ref{sec:algcrit}.
The duality symmetry~(\ref{jdusig}) implies the relation
\beq
C^\pm(v)C^\mp(-v)=1.
\label{jduc}
\eeq

\subsection{Convergent regime: $v>\V$ or $p>p_c$}

In this regime, the drift velocity $\V$ of the random walker
belongs to the domain in which the mean velocities $V_n$ are constrained.
These constraints are therefore less and less stringent as time goes on.
Indeed $P^\pm(n,v)$ go to unity exponentially fast, with $1-P^\pm(n,v)$
given by a large-deviation expression [see eq.~(\ref{jregp3})].

The Sparre Andersen relation~(\ref{jic}) now leads to the estimate
\beq
f^\pm(z,v)\approx\frac{F^\pm(v)}{1-z}\qquad(z\to1).
\eeq
In other words, the survival probabilities converge toward
non-trivial limits for infinite times:
\beq
F^\pm(v)=\lim_{n\to\infty}F^\pm(n,v),
\label{jfcv}
\eeq
hence the word `convergent regime'.
The limit survival probabilities have the formal expressions
\beq
F^\pm(v)=\exp\left(-\sum_{n\ge1}\frac{1-P^\pm(n,v)}{n}\right)\qquad(v>\V).
\label{jf}
\eeq

The limit probabilities $F^\pm(v)$ have in general a very rich dependence
on $v$ and $p$, which is not apparent in the formal expressions~(\ref{jf}),
and which will be investigated in the following sections.
For the time being, we want to underline the existence
of a universal relationship
between the $p$-dependence of $F^\pm(v)$ as $p\to p_c+0$
and the amplitudes $C^\pm(v)$ corresponding to the marginal regime $(p=p_c)$.
For $p-p_c$ small and positive,
the one-time distribution functions $P^\pm(n,v)$
can be approximated by their values at $p=p_c$ as long as $n(p-p_c)\ll1$,
while they can be estimated from the central limit theorem for $n(p-p_c)\sim1$:
\beq
P^\pm(n,v)\approx\frac{1}{2}
\left[1+\mathrm{erf}\left(\frac{n(p-p_c)}{2^{3/2}p_cq_c}\right)\right],
\eeq
where erf is the error function.
By inserting the above estimate into eq.~(\ref{jf}),
and splitting the sum over $n$ into three sums,
respectively corresponding to the ranges
$1\le n\le n_0$, $n_0<n\le n_1$, and $n>n_1$,
with $1\ll n_0\ll 1/(p-p_c)\ll n_1$, we obtain after some manipulations
\beqa
F^\pm(v)
&\approx&C^\pm(v)\left(\frac{2}{p_cq_c}\right)^{1/2}(p-p_c)\nonumber\\
&\approx&C^\pm(v)\left(\frac{2}{1-v^2}\right)^{1/2}(v-\V).
\label{jfpc}
\eeqa
The survival probabilities $F^\pm(v)$ thus vanish linearly as $p\to p_c$,
with amplitudes proportional to the constants $C^\pm(v)$,
characteristic of the fall off~(\ref{jfc}) in the marginal regime $(p=p_c)$.

The quantities introduced to describe the asymptotic long-time behavior
of the survival probabilities $F^\pm(n,v)$ in the three different regimes,
namely the periodic functions $b^\pm(x,v)$, the constants $C^\pm(v)$,
and the limit survival probabilities $F^\pm(v)$,
bear a highly non-trivial dependence on the parameters $p$ and $v$,
which is hidden in their formal expressions~(\ref{jb}),
(\ref{jc}), and~(\ref{jf}).
The rest of this paper is devoted to an investigation of all these quantities,
including their dependence on $p$ and $v$.
Throughout the following, when the context permits, we shall suppress the
explicit dependence on $v$, writing e.g. $F^\pm$ for $F^\pm(v)$.

\section{Basic concepts}
\label{sec:prel}

\subsection{The survival probability as a distribution function}
\label{sec:properties}

Define the maximal velocity $V\max$ of a given (infinite) random walk
as the supremum of all the instantaneous mean velocities $V_n$:
\beq
V\max=\sup_n V_n.
\eeq
This quantity is a random variable,
as it depends on the walk under consideration.
The survival probabilities $F^\pm(v)$ can be interpreted
as the distribution functions of this random variable,
for a fixed value of $p$.
We have indeed
\beq
F^\m(v)=\prob\big\{V\max<v\big\},\qquad F^\p(v)=\prob\big\{V\max\le v\big\}.
\eeq

With probability one the random variable $V\max$ is rational,
and it lies in the range $\V<V\max\le1$.
These properties originate in the following two facts.
The $V_n$ converge with probability one to the drift velocity $\V$,
and the differences $V_n-\V$ take both signs.
As a consequence, with probability one there is some finite $n$ such that
$V\max=V_n$.
We have therefore
\beq
F^\pm(v)=0\quad(v\le\V),\qquad F^\pm(v)=1\quad(v>1).
\eeq
The lower edge $(v=\V)$ corresponds to the marginal case
studied in sec.~\ref{sec:crit},
while the upper edge $(v=1)$ will be investigated in sec.~\ref{sec:vun}.

Whenever the slope $v$ is irrational, we have $F^\p(v)=F^\m(v)$,
and the distribution function is continuous.
On the contrary, when $v$ is rational, quantities with superscripts
$\p$ and $\m$ are different from each other in general.
The discontinuity $\Pi(v)$ of the distribution function
is nothing but the probability
that the maximal velocity $V\max$ assumes the value $v$:
\beq
\Pi(v)=F^\p(v)-F^\m(v)=\prob\big\{V\max=v\big\},
\eeq
and we have
\beq
F^\p(v)=\lim_{w\to v+0}F(w),\qquad F^\m(v)=\lim_{w\to v-0}F(w).
\eeq

The above quantities can be related as follows.
Let $Q(v)$ be the probability that the random walker
makes its first step to the left and then touches the wall at least once:
\beq
Q(v)=\sum_{n\ge2}
\prob\big\{V_m<v\,\hbox{ for }\,m=1,\dots,n-1\,\hbox{ and }\,V_n=v\big\}.
\label{jqdef}
\eeq
This quantity is non-zero for any rational $v$ in the range $-1<v\le1$.
We can split any walk contributing to $Q(v)$ into two independent walks:
a finite walk from the origin to the first coincidence time $n$
such that $V_n=v$, and the rest, which is an infinite walk.
Consider a walk contributing to $F^\p(v)$.
It either contributes to $F^\m(v)$ or to $\Pi(v)$.
In the second case, we have $V\max=v$,
and with probability one there is some $n$
such that $V_n=v$ for the first time.
The complete walk splits into two independent walks:
a finite walk from the origin to the first coincidence,
and the rest, which is an infinite walk contributing to $F^\p(v)$.
So, in order to go from $Q(v)$ to the difference $F^\p(v)-F^\m(v)$,
one just replaces the arbitrary infinite part
by a walk contributing to $F^\p(v)$.
In other words, we have
\beq
\Pi(v)=F^\p(v)-F^\m(v)=Q(v)F^\p(v),
\label{jpi}
\eeq
with this quantity vanishing for $v\le\V$ and being positive for $v>\V$,
or else
\beq
F^\m(v)=\bigl(1-Q(v)\bigr)F^\p(v).
\label{jq}
\eeq

The case of a rational slope $v$ will be investigated in sec.~\ref{sec:arb}.
It turns out that some general combinatorial statements
can be conveniently presented by keeping $v$ arbitrary.
This is the purpose of the rest of sec.~\ref{sec:prel}.

\subsection{Truncated Pascal triangle}
\label{sec:kincomb}

The oriented lattice edges crossing the wall from left to right will
play an important role in what follows.
Indeed the walks which contribute to the survival probability,
i.e., those which remain on the left forever,
are those which start to the left and never pass through a crossing edge.

We assume for a while that $v$ is irrational.
As shown in Figure~5, we index the crossing edges by the integer $k\ge1$,
i.e., by the number of steps to the left,
defined in eq.~(\ref{jdefk}).
The endpoint of the $k$-th crossing edge has coordinates $(n,x)=(n_k,n_k-2k)$,
where $n_k$ is defined by the inequalities $k/p_c<n_k<k/p_c+1$, or equivalently
\beq
n_k=1+\Int(k/p_c).
\label{jdefnk}
\eeq

When $v$ is rational, we have to be more careful and define two sequences.
The edges corresponding to $F^\m(v)$ may end but not start on the wall,
while those corresponding to $F^\p(v)$ may start but not end on the wall,
hence the prescriptions
\beq
n_k^\m=1+\Int^\p(k/p_c),\qquad n_k^\p=1+\Int^\m(k/p_c).
\label{jnpm}
\eeq
Hence, if $v$ is rational, any construction involving the
sequence of crossing edges will give two different outputs in general,
depending on which family of crossing edges is considered.
In this case, we may, and shall sometimes, start the sequence of crossing
edges with $k=0$.

$$\centerline{\epsfbox{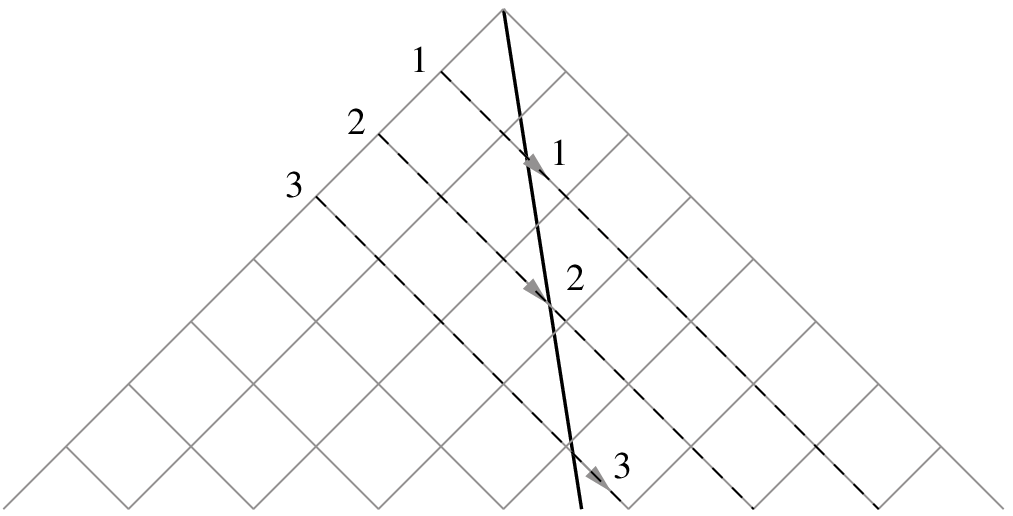}}$$
\capt{5}{M4}{Labeling of crossing edges by an integer $k=1$, 2, 3,
equal to the number of steps of the walker to the left.}

Let us emphasize that our problem
can be recast as a purely combinatorial one.
We assume again for simplicity that $v$ is irrational.
Let $\N_{n,k}$ be the number of $n$-step walks starting from the origin,
remaining on the left of the wall,
and ending at the point $(n,n-2k)$, with the notations~(\ref{jdefk}).
The numbers $\N_{n,k}$ obey a recursion relation
very similar to that obeyed by the binomial coefficients $\bin{n}{k}$:
\beq
\N_{n,k}=\left\{\matrix{
\N_{n-1,k-1}+\N_{n-1,k}\hfill&(k>np_c),\hfill\cr
0\hfill&(k<np_c),\hfill
}\right.%}
\label{ntx}
\eeq
with the initial condition $\N_{0,0}=1$, hence in particular $\N_{n,n}=1$.
The recursion~(\ref{ntx}) leads to a double array of integers,
shown in Figure~6, that we refer to as a truncated Pascal triangle.

If the slope $v$ is rational, i.e., $p_c$ is rational,
one has to consider two series of integers, $\N^\pm_{n,k}$,
which obey the recursion relations
\beqa
\N^\p_{n,k}&=&\left\{\matrix{
\N^\p_{n-1,k-1}+\N^\p_{n-1,k}\hfill&(k\ge np_c),\hfill\cr
0\hfill&(k<np_c),\hfill}\right.%}
\nonumber\\
\N^\m_{n,k}&=&\left\{\matrix{
\N^\m_{n-1,k-1}+\N^\m_{n-1,k}\hfill&(k>np_c),\hfill\cr
0\hfill&(k\le np_c),\hfill
}\right.%}
\label{ntrat}
\eeqa
again with $\N^\pm_{0,0}=1$.

As the weight of a finite walk only depends on its endpoint,
the integers $\N_{n,k}$ formally contain all the relevant information
for the computation of the survival probability.
More powerful techniques allowing to extract this information
will be exposed in the following.

$$\centerline{\epsfbox{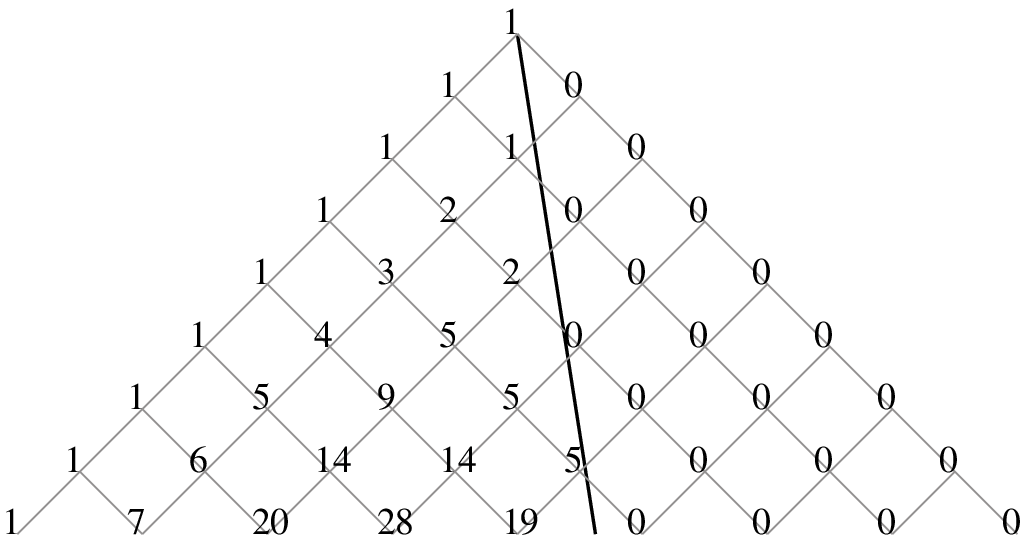}}$$
\capt{6}{M5}{Combinatorial approach: truncated Pascal triangle.}

\subsection{Probability flow equation}
\label{sec:jpfe}

The recursion relations~(\ref{ntx}),~(\ref{ntrat})
for the integers $\N^\pm_{n,k}$
are different from those for the binomial coefficients at the crossing edges.
This indicates that there is a probability flow through the wall.
To analyze this probability flow, we define the numbers $A^\pm_k$
as the elements of the truncated Pascal triangle
at the beginning of a crossing edge:
\beq
A^\pm_k=\N^\pm_{n_k-1,k}.
\label{jan}
\eeq

Eqs.~(\ref{ntx}),~(\ref{ntrat}) imply that $A^\pm_k$ walks are absorbed
at the $k$-th crossing edge.
In order to count them with the right probability
[see eqs.~(\ref{jcn1}),~(\ref{claude}) below],
we have to put these numbers at the end of the crossing edges,
obtaining thus the picture shown in Figure~7.
In the above example, we get the sequence $1,2,5,19,\dots$

This construction leads to the most important (though elementary)
equation of this paper.
The probability to begin with a left step is $p$,
while $A^\pm_k$ walks are lost at the $k$-th crossing edge,
each with a probability $p^kq^{n^\pm_k-k}$,
hence the {\it probability flow equation}
\beq
F^\pm(n,v)=p-\sum_{k=1}^{\Int^\pm(np_c)}A^\pm_k p^kq^{n^\pm_k-k},
\label{jcn1}
\eeq
where we have used the equivalence between the inequalities
$n^\pm_k\le n$ and $k\le\Int^\pm(np_c)$.
Taking the $n\to\infty$ limit, we obtain the following flow equation
\beq
F^\pm=p-\sum_{k\ge1}A^\pm_k p^kq^{n^\pm_k-k}
\label{claude}
\eeq
for the limit survival probabilities $F^\pm$,
which still depend on $p$ and $v$.

$$\centerline{\epsfbox{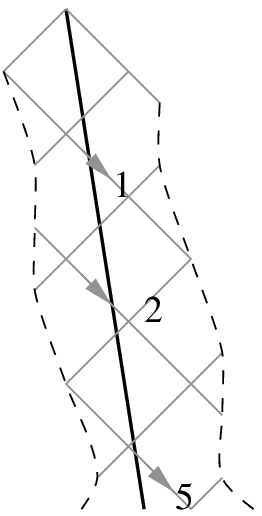}}$$
\capt{7}{M6}{The probability flow is characterized by integers $A^\pm_k$,
living on the crossing edges [cf. Figure~5].
In the example, we have $A^\pm_1=1$, $A^\pm_2=2$, $A^\pm_3=5$, and so on.}

The flow equations~(\ref{jcn1}) and~(\ref{claude})
will be used several times in the following.
In particular, eq.~(\ref{claude}) will be used in sec.~4
in order to derive the second and third recursion relations
satisfied by the sequences $A^\pm_k$.
Before we turn to these matters,
we can obtain some information on the analytic structure of $F^\pm$ in $p$.
Eq.~(\ref{claude}) shows that the expression
\beq
\sum_{k\ge1}A^\pm_kp^kq^{n^\pm_k-k}
\label{jsum}
\eeq
with $q=1-p$,
is a convergent sum of positive numbers for any $p$ in the interval $[0,1]$.
Using the definition~(\ref{jdefnk}), we have
\beq
\label{splitpt}
p^kq^{n^\pm_k-k}=z(p)^kq^{1-\Frac^\mp(k/p_c)},
\eeq
with
\beq
z(p)=p\,q^{q_c/p_c}.
\label{jdefz}
\eeq
As the second factor $q^{1-\Frac^\mp(k/p_c)}$ is bounded,
we can concentrate on the first factor $z(p)^k$.
The maximum of the function $z(p)$ for $p$ in $[0,1]$
is reached for $p=p_c$.
Hence the sum~(\ref{jsum}) is a holomorphic function of $p$
in a neighborhood of the origin,
and it can be expanded term by term as long as $\abs{z(p)}<z(p_c)$.
Suppose now $p>p_c$.
Eq.~(\ref{splitpt}) implies $p^kq^{n^\pm_k-k}<p_c^kq_c^{n^\pm_k-k}$
for each $k$, so that
\beq
F^\pm>p-\sum_{k\ge1}A^\pm_k p_c^kq_c^{n^\pm_k-k}=p-p_c.
\eeq
This property gives a proof of the fact, announced before,
that $F^\pm>0$ for $p>p_c$.
This also shows that the sum~(\ref{jsum}) is not analytic at $p=p_c$,
because if it were so the equality $F^\pm=0$,
which holds for $p\le p_c$, could be continued across $p=p_c$.

\section{Arbitrary slopes: recursion relations}
\label{sec:arb}

This section is devoted to the investigation of the sequences
of integers $A^\pm_k$.
The slope $v$ is an arbitrary number throughout this section,
so that we shall omit the superscripts $\pm$, for brevity.

\subsection{Continuity of the path}

By expressing the continuity of the path of the random walker,
we shall obtain the first recursion relation defining the integers $A_k$.

Consider a walk of length $n_k$ that starts to the left
and terminates at the end of the $k$-th crossing edge.
There are $\bin{n_k-1}{k-1}$ such walks.
Either this walk crosses the wall for the first time along the
$k$-th crossing edge (there are $A_k$ such walks),
or it has crossed the wall for the first time along some earlier crossing edge,
say the $\ell$-th one.
In the second case, the full walk is the concatenation of one of the $A_\ell$
walks that cross for the first time along the $\ell$-th crossing edge,
and of any walk from $(n_\ell,\ell)$ to $(n_k,k)$
(there are $\bin{n_k-n_\ell}{k-\ell}$ such walks).
We thus obtain our first recursion relation for the sequence
of integers $A_k$:
\beq
\label{rec1}
\bin{n_k-1}{k-1}=\sum_{\ell=1}^k\bin{n_k-n_\ell}{k-\ell}A_\ell
=A_k+\sum_{\ell=1}^{k-1}\bin{n_k-n_\ell}{k-\ell}A_\ell.
\eeq
Eq.~(\ref{rec1}) clearly leads to integer values for $A_k$,
but not obviously to positive ones.
The positivity of the $A_k$ can be checked on the first few of them, i.e.,
\beq
A_1=1,\quad A_2=n_1-1,\quad A_3=\frac{1}{2}(n_1-1)(2n_2-n_1-2),\quad\hbox{etc.}
\eeq

The recursion relation~(\ref{rec1}) looks like a convolution.
It fails however to be an exact convolution,
because $n_k-n_\ell\ne n_{k-\ell}$ in general.
Nevertheless the difference between these integers is small
(zero or one in absolute value).

\subsection{Probability flow}

The second recursion relation for the integers $A_k$
relies on the probability flow equation~(\ref{claude}),
and on the analyticity at small $p$ of the sum~(\ref{jsum}),
proved at the end of sec.~\ref{sec:jpfe}.
We know that $F=0$ for fixed $v$ and small enough $p$.
Hence we can apply contour integrals to both sides of the relation
\beq
p=\sum_{k\ge1}A_k p^kq^{n_k-k}.
\label{jpflow}
\eeq
To be more precise, let $\ell\ge1$ be an integer,
and $c_\ell(p)$ be a holomorphic function in a neighborhood of the origin,
with a Taylor expansion starting as $c_\ell(p)=p^{\ell+1}+\cdots$
Integrating along a small contour around the origin, we have
\beq
\label{genrec}
\oint\frac{\d p}{2\pi\i c_\ell(p)}p=\sum_{k\ge1}
A_k\oint\frac{\d p}{2\pi\i c_\ell(p)}\,p^kq^{n_k-k}=A_\ell+
\sum_{k=1}^{\ell-1}A_k\oint\frac{\d p}{2\pi\i c_\ell(p)}\,p^kq^{n_k-k}.
\eeq
Indeed the integral in the two rightmost sides equals 0 for $k>\ell$,
and 1 for $k=\ell$.

We can take any sequence of functions $c_\ell(p)$,
and get a corresponding recursion relation which determines the $A_k$.
The following three cases will prove useful in the sequel.

\begin{itemize}

\item{
The most obvious candidate is
\beq
c_\ell(p)=p^{\ell+1},
\eeq
which leads to our second recursive definition of the $A_k$:
\beq
\label{rec2}
\delta_{k,1}=A_k+\sum_{\ell\le k-1,\ k\le n_\ell}
(-1)^{k-\ell}\bin{n_\ell-\ell}{k-\ell}A_\ell,
\eeq
where $\delta_{k,\ell}$ is the Kronecker symbol.
Again, the integrality of the $A_k$ is obvious, but their positivity is not.
}

\item{
Another simple choice reads
\beq
c_\ell(p)=p^{\ell+1}q^{n_\ell-\ell+1}.
\eeq
Surprisingly enough, we recover our first
recursion formula~(\ref{rec1}), expressing the continuity of the path.
This example illustrates the generality of eq.~(\ref{genrec}).
}

\item{
Let us come back to eq.~(\ref{splitpt}),
which shows that the sum~(\ref{jsum})
is almost an entire series in $z(p)$, with the definition~(\ref{jdefz}),
up to a small positive power of $q$, with an exponent in the range $[0,1]$.
If we now divide expression~(\ref{jsum}) by $q$, we have again almost
an entire series in $z(p)$,
up to a small negative power of $q$, with an exponent in the range $[-1,0]$.
It is therefore natural to evaluate the contour integrals
of eq.~(\ref{genrec}) with the weights
\beq
\frac{\d p}{c_\ell(p)}=\frac{\d z}{z^{\ell+1}}\quad\hbox{and}\quad
\frac{\d p}{c_\ell(p)}=\frac{\d z}{z^{\ell+1}(1-p(z))},
\eeq
where $p(z)$ is the inverse function of $z(p)$, defined in eq.~(\ref{jdefz}).
After somewhat lengthy computations, which boil down to binomial expansions,
we respectively obtain
\beqa
\frac{\Gamma(k/p_c-1)}{k!\,\Gamma(kq_c/p_c)}
&=&A_k-\sum_{\ell=1}^{k-1}
\frac{\Gamma(k/p_c-n_\ell)}{(k-\ell)!\,\Gamma(kq_c/p_c+1-n_\ell+\ell)}
\,\bigl(1-\Frac(\ell/p_c)\bigr)A_\ell,\nonumber\\
\frac{\Gamma(k/p_c+1)}{k!\,\Gamma(kq_c/p_c+2)}
&=&A_k+\sum_{\ell=1}^{k-1}
\frac{\Gamma(k/p_c-n_\ell+1)}{(k-\ell)!\,\Gamma(kq_c/p_c+2-n_\ell+\ell)}
\,\Frac(\ell/p_c)\,A_\ell.
\label{jrecd}
\eeqa
}
\end{itemize}

The recursion relations~(\ref{jrecd}) seem rather complicated,
so that even the integrality of the $A_k$ is not obvious.
Eqs.~(\ref{jrecd}) are nevertheless of interest.
Indeed all the coefficients in the sums have the same sign,
since the $A_k$ are positive.
We thus obtain the following bounds
\beq
\label{bounds}
\frac{\Gamma(k/p_c-1)}{k!\,\Gamma(kq_c/p_c)}
\le A_k\le\frac{\Gamma(k/p_c+1)}{k!\,\Gamma(kq_c/p_c+2)}.
\eeq

The case where $p_c=1/N$, with $N$ an integer, is of special interest.
In this situation, $\ell/p_c=\ell N$ is an integer for any $\ell$.
Re-introducing the $\pm$ superscripts for a while,
we thus have $\Frac^\p(\ell/p_c)=1$ if we are interested in $F^\m$,
and $\Frac^\m(\ell/p_c)=0$ if we are interested in $F^\p$.
As a consequence, the $A_k^\m$ saturate the lower bound of eq.~(\ref{bounds}),
while the $A_k^\p$ saturate the upper bound.
We shall recover these properties in sec.~\ref{sec:vemun}
with more powerful techniques.

In the general case, we set
\beq
A^\pm_k=B^\pm_k\frac{\Gamma(k/p_c)}{k!\,\Gamma(kq_c/p_c+1)}.
\label{jdefb}
\eeq
Eq.~(\ref{bounds}) shows that the prefactors $B^\pm_k$ are bounded.
In other words, using again Stirling's formula, we obtain the estimates
\beq
A^\pm_k\approx B^\pm_k\,\frac{p_c}{(2\pi q_c\,k^3)^{1/2}}
\,\left(p_c\,q_c^{q_c/p_c}\right)^{-k}\qquad(k\gg1).
\label{jasya}
\eeq
The prefactors $B^\pm_k$ obey the simple bounds
\beq
q_c\le B^\pm_k\le\frac{1}{q_c}.
\label{bbounds}
\eeq
We shall come back in sec.~\ref{sec:algcrit}
to these prefactors, which are non-trivial in general.

\subsection{Duality}
\label{sec:du}

The recursion relations derived in the previous section
are based on the fact that $F=0$ for small enough $p$.
We now aim at using the vanishing of $F$ on the full interval $0\le p\le p_c$.
We call the following approach duality, as in the beginning
of sec.~\ref{sec:sa},
since it relates the survival probability in presence of the obstacle
with slope $v$ to that with slope $-v$.
In sec.~\ref{sec:algtrickdual} we shall
give a simple combinatorial argument explaining why the slopes $v$ and~$-v$,
i.e., $p_c$ and $q_c$, are related, at least for rational $v$.

We start again from the identity~(\ref{jpflow}),
\beq
p-\sum_{k\ge1}A_kp^kq^{n_k-k}=0\qquad(0\le p\le p_c),
\label{jidp}
\eeq
and the similar equation for the slope $-v$, with integers $\w A_k$,
\beq
p-\sum_{k\ge1}\w A_kp^kq^{\w n_k-k}=0\qquad(0\le p\le q_c).
\label{jidpp}
\eeq
By substituting $q=1-p$ for $p$ in this last equation, we get
\beq
q-\sum_{k\ge1}\w A_kq^kp^{\w n_k-k}=0\qquad(p_c\le p\le 1).
\label{jidq}
\eeq
Multiplying eqs.~(\ref{jidp}),~(\ref{jidq}), we obtain
\beq
\label{dual}
\left(p-\sum_{k\ge1}A_kp^kq^{n_k-k}\right)
\left(q-\sum_{k\ge1}\w A_kq^kp^{\w n_k-k}\right)=0\qquad(0\le p\le1).
\eeq

For the rest of this section, we assume that $v$ is irrational,
while the case where $v$ is rational is treated
in sec.~\ref{sec:algtrickdual}.
For $v$ irrational and $k\ge1$, the integer $n_k$ corresponding to the
$k$-th crossing with the wall at slope $v$ is defined by the inequalities
$0<p_cn_k-k<p_c$, while the integer $\w n_k$ corresponding to the
$k$-th crossing with the wall at slope $-v$ is defined by the inequalities
$-q_c<p_c\w n_k-(\w n_k-k)<0$.
Hence the fractional parts of $p_cn_k$ and $p_c\w n_k$
live in the disjoint intervals $[0, p_c]$ and $[-q_c, 0]$.
Consequently, the set of the $n_k$ and the set of the $\w n_k$ have no
integer in common.
Note that $n_k$ and $\w n_k$ are both larger than 1.

Conversely, for $n>1$, there is a unique integer $k_n$
such that $-q_c<np_c-k_n<p_c$.
We set $\theta_n=np_c-k_n$.
Equivalently, with the definitions~(\ref{jdefif}), we have
\beq
k_n=1+\Int\left((n-1)p_c\right),\qquad\theta_n=-q_c+\Frac\bigl((n-1)p_c\bigr).
\label{jkn}
\eeq

Let $I^+$ be the set of integers $n>1$ such that $\theta_n>0$.
For such an $n$, there is an integer $k\ge1$ such that $n=n_k$,
namely $k=k_n$, and we have
\beq
n\in I^+:\quad\theta_n=p_c\left(1-\Frac\left(\frac{k_n}{p_c}\right)\right).
\label{jthplus}
\eeq
Similarly, let $I^-$ be the set of integers $n>1$ such that $\theta_n<0$.
For such an $n$, there is an integer $k\ge1$ such that $n=\w n_k$,
namely $k=n-k_n$, and we have
\beq
n\in I^-:\quad\theta_n=-q_c\left(1-\Frac\left(\frac{n-k_n}{q_c}\right)\right).
\label{jthmoins}
\eeq
The above two sets are a partition of the integers $n>1$.
We can therefore define a sequence of integers $\oA_n$ for $n>1$ by
\beq
\oA_n=\left\{\matrix{
A_{k_n}\hfill&\hbox{if}\,\,n\in I^+,\cr
\w A_{n-k_n}\hfill&\hbox{if}\,\,n\in I^-.
}\right.%}
\label{jdefan}
\eeq

A last geometrical observation is in order.
It follows from the definition of the sequence $k_n$ that
$k_n\le k_{n+1}\le k_n+1$.
Hence the sequence
\beq
X_n=n-2k_n
\label{jdefxn}
\eeq
defines a canonical walk, shown in Figure~8.

This walk is closest to the wall in a well-defined sense:
the point $(n,n-2k_n)$ is the endpoint of an edge crossing the wall,
either from left to right or from right to left.
Equivalently, the next step of the canonical walk is to the left
if it starts to the right of the wall, and vice versa.
We identify the symbol $X_n$ with the polynomial
\beq
X_n=p^{k_n}q^{n-k_n},
\label{jdefx}
\eeq
whose total degree is $n$, and whose difference
of partial degrees in $p$ and $q$
gives the position $X_n$, as in eq.~(\ref{jdefxn}).

$$\centerline{\epsfbox{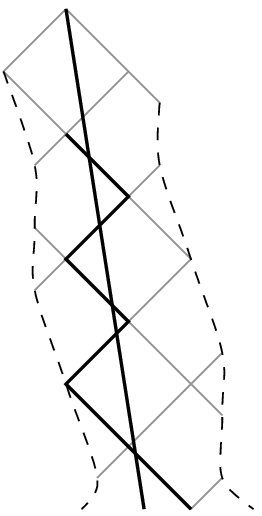}}$$
\capt{8}{M7}{The canonical walk.}

With the notations~(\ref{jdefan}),~(\ref{jdefx}),
eq.~(\ref{dual}) can be recast as
\beq
\left(1-q-\sum_{n\in I^+}\oA_nX_n\right)
\left(q-\sum_{n\in I^-}\oA_nX_n\right)=0\qquad(0\le p\le1).
\eeq
Let us expand the product.
It is easy to check that if $n$ and $m$
have different $I$-signs, then $X_nX_m=X_{n+m}$.
Moreover, if $n\in I^+$, $pX_n=X_{n+1}$, and if $n\in I^-$, $qX_n=X_{n+1}$.
Finally, $pq=X_2$.
We set $\oA_1=1$, and $I_0^+=I^+\cup\{1\}$ and $I_0^-=I^-\cup\{1\}$.
We obtain after some algebra
\beq
\sum_{n>1}\oA_nX_n=\sum_{m\in I_0^-,\ n\in I_0^+}\oA_m\oA_nX_{n+m}
\qquad(0\le p\le1).
\eeq
A heuristic term-by-term identification leads to a third
recursive definition of the integers $\oA_n$:
\beq
\oA_\ell=\sum\dbli{m\in I_0^-,\ n\in
I_0^+}{m+n=\ell}\oA_m\oA_n\qquad(\ell\ge2).
\label{rec3}
\eeq
An algebraic proof of this relation
will be given in sec.~\ref{sec:algtrickdual}.

\section{Rational slopes: algebraic approach}
\label{sec:algtrick}

In this section we investigate in further detail the outcomes
of the techniques exposed in sec.~\ref{sec:arb},
in the case where $v$ is rational.
Algebraic functions play a central role in this analysis.

We write a rational slope in terms of two relatively prime integers
$M,\w M\ge1$ as
\beq
N=M+\w M,\quad v=\frac{\w M-M}{N},
\quad p_c=\frac{M}{N},\quad q_c=\frac{\w M}{N}.
\label{jdefmn}
\eeq
The integers $n^\pm_k$, defined in eq.~(\ref{jnpm}), are skew periodic,
in the sense that
\beq
n^\pm_{k+M}=n^\pm_k+N,
\label{jnkper}
\eeq
and we have
\beq
n^\m_k=n^\p_k\quad(k=1,\dots,M-1),\qquad n^\m_M=N,\quad n^\p_M=N+1.
\label{jnklim}
\eeq
We recall that, for each rational slope $v$,
we have to consider two different sequences of integers, $A_k^\pm$,
associated with the survival probabilities $F^\pm$.

\subsection{Continuity of the path}
\label{sec:algcont}

As a consequence of the property~(\ref{jnkper}) of the integers $n^\pm_k$,
the recursion relation~(\ref{rec1}) can be split into $M$ blocks,
one for each value of $k$ modulo $M$.

We introduce the following formal power series in $t$:
\beqa
&&F^\pm_k(t)=\sum_{K\ge0}A^\pm_{KM+k}t^{K}\qquad(k=1,\dots,M),
\label{defF}\\
&&G^\pm_k(t)=\sum_{K\ge0}\bin{KN+n^\pm_k-1}{KM+k-1}t^{K}\qquad(k=1,\dots,M),
\label{defG}\\
&&\!\!G^\pm_{k,\ell}(t)
=\sum\dbli{K\ge0}{KM+k-\ell\ge0}\bin{KN+n^\pm_k-n^\pm_\ell}{KM+k-\ell}t^{K}
\qquad(k,\ell=1,\dots,M),
\label{defGG}
\eeqa

Eq.~(\ref{rec1}) becomes a set of $M$ coupled linear equations, of the form
\beq
\label{G=GF}
\sum_{\ell=1}^{M}G^\pm_{k,\ell}(t)F^\pm_\ell(t)
=G^\pm_k(t)\qquad(k=1,\dots,M).
\eeq
The solution of this linear system yields the functions $F^\pm_k(t)$,
which are generating functions for the integers $A^\pm_k$.

So far $t$ is a formal expansion variable.
There is however a natural choice for it, namely
\beq
t=p^Mq^{\w M}=p^M(1-p)^{N-M}=z(p)^M,
\label{jdeft}
\eeq
so that the survival probabilities $F^\pm$
can be expressed in terms of the functions $F^\pm_k(t)$.
Indeed, with the definitions~(\ref{jdeft}) of the variable $t$
and~(\ref{defF}) of the functions $F^\pm_k(t)$,
the probability flow equation~(\ref{claude}) can be recast as
\beq
F^\pm=p-\sum_{k=1}^Mp^kq^{n^\pm_k-k}F^\pm_k(t).
\label{clauderat}
\eeq

We now show that the functions $G^\pm_k(t)$ and $G^\pm_{k,\ell}(t)$
are algebraic functions of $t$.
These functions are special cases of the functions
\beq
H_{ij}(t)=\sum_{K\ge0}\bin{KN+j}{KM+i}t^K,
\label{jhdef}
\eeq
for $i,j$ non-negative integers.
Our starting point is the following contour integral representation
of the binomial coefficient:
\beq
\bin{I}{J}=\oint\frac{\d z}{2\pi\i}\,\frac{(1+z)^I}{z^{J+1}}
=\oint\frac{\d u}{2\pi\i}\,\frac{1}{u^{J+1}(1-u)^{I-J+1}},
\label{jintij}
\eeq
where the integral is along a small contour around the origin.
The second expression is obtained by means of the change of variable
$z=u/(1-u)$.
Applying eq.~(\ref{jintij}) to $(I,J)=(KN+j, KM+i)$,
and summing over $K$, we get
\beq
H_{ij}(t)=\oint\frac{\d u}{2\pi\i}\,
\frac{u^{M-i-1}(1-u)^{N-M+i-j-1}}{u^M(1-u)^{N-M}-t}.
\label{jhint}
\eeq

Now, we need to know which roots of the denominator
are inside the integration contour.
The geometric sum over $K$ is convergent for $\abs{t}<\abs{u^M(1-u)^{N-M}}$.
On the other hand, the polynomial equation
\beq
u^M(1-u)^{N-M}=t
\label{jupoly}
\eeq
has $M$ `small' roots $u_\alpha$, with $\alpha=1,\dots,M$, such that
\beq
u_\alpha\approx t^{1/M}\o^{\alpha-1}\qquad(t\to0),
\label{judef}
\eeq
with
\beq
\o=\e^{2\pi\i/M}.
\label{jdefo}
\eeq
The small roots can be continued in $t$.
For $t$ real in the range $0\le t\le t_c$, with
\beq
t_c=p_c^Mq_c^{N-M}=\frac{M^M(N-M)^{N-M}}{N^N},
\eeq
the $M$ small roots remain inside the circle $\abs{u}=p_c$,
while the other $N-M$ roots of eq.~(\ref{jupoly}),
the `large' ones, lie inside the circle $\abs{1-u}=q_c$.
The two circles in the $u$-plane are tangent to each other at $u=p_c$.
For $p<p_c$, one of the small roots is $p$ itself,
while for $p>p_c$, one of the large roots is $p$ itself.
Figure~9 shows a plot of the roots of eq.~(\ref{jupoly}), for $M=4$, $N=11$,
and $p=1/2>p_c=4/11$, i.e., $t=1/2048$, while $t_c=108/3125$.
The root $p=1/2$ is shown as a large empty symbol.

\vskip 8.2cm{\hskip 1.2cm}
\includegraphics{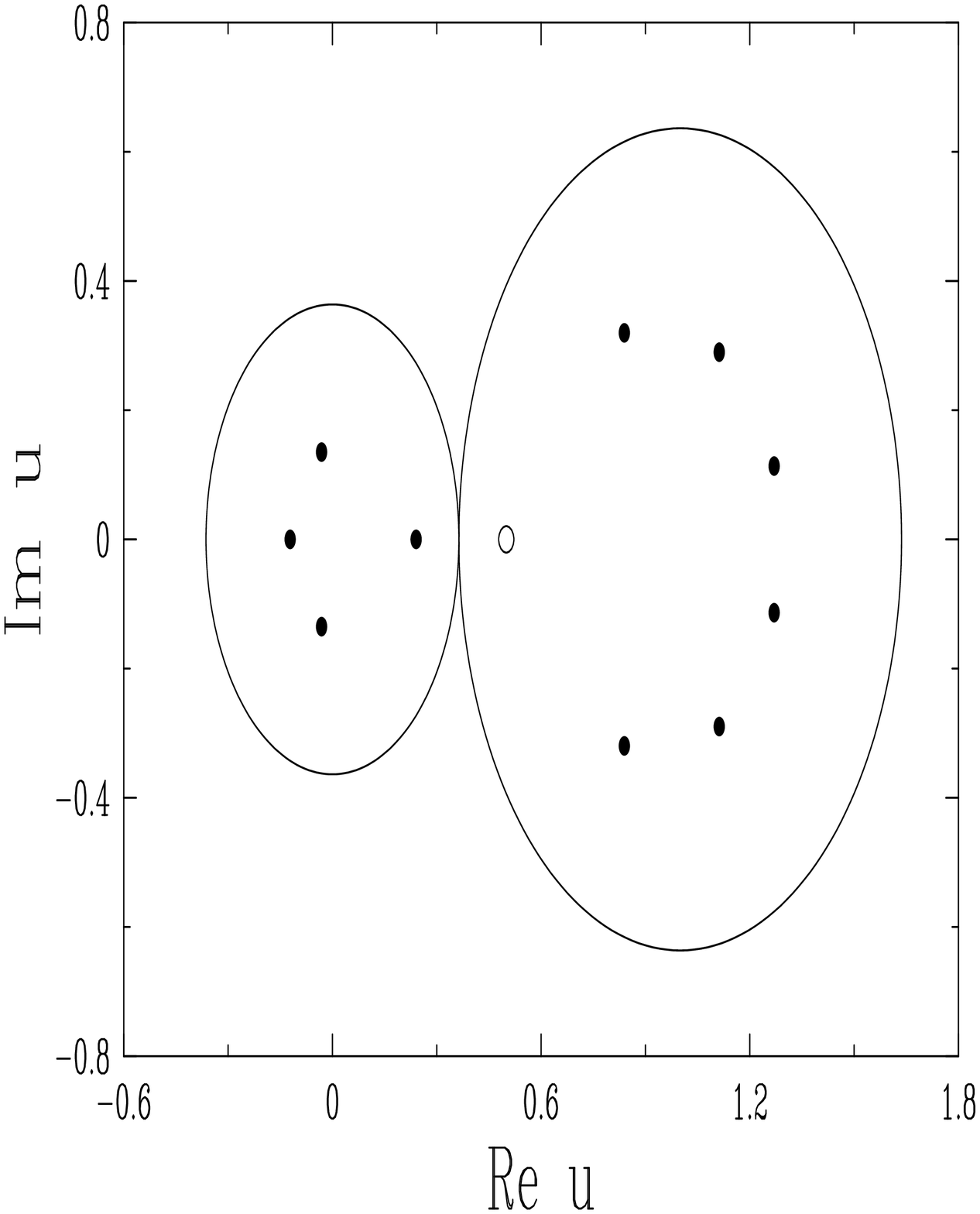}

\capt{9}{J1}{Roots of the polynomial equation~(\ref{jupoly})
for $M=4$, $N=11$, and $p=1/2$.}

For $\abs{t}$ small enough,
the contour of eq.~(\ref{jhint}) contains exactly the $M$ small
roots $u_\alpha$, so that we have
\beq
H_{ij}(t)=\sum_{\alpha=1}^M
\frac{1}{(M-Nu_\alpha)u_\alpha^i(1-u_\alpha)^{j-i}}.
\eeq
Each of the $u_\alpha(t)$ is an algebraic function of degree $N$,
since it obeys eq.~(\ref{jupoly}).
As a consequence, $H_{ij}(t)$ is an algebraic function of degree $\bin{N}{M}$,
for any non-negative integers $i,j$.
Indeed, choosing one branch of the function $H_{ij}(t)$
amounts to choosing $M$ branches of $u_\alpha(t)$ among $N$.

The algebraic function $H_{ij}(t)$ is singular when two
roots of the polynomial equation~(\ref{jupoly}) coincide,
in such a way that the contour of eq.~(\ref{jhint}) gets pinched.
The only non-trivial singularity corresponds to $t=t_c$,
where the real positive small root $u=u_1$ merges with the root $u=p$
into a double root at $u=p_c$.
In the vicinity of this point, we introduce the notation
\beq
p=p_c+\delta p,\qquad t\approx t_c\left(1-\frac{N}{2p_cq_c}(\delta p)^2\right).
\label{jcrit}
\eeq
For $\delta p>0$, the first root $u_1$ has a correction in $\delta p$,
while the other $M-1$ roots are regular in $t$,
so that their corrections are of order $(\delta p)^2$:
\beq
u_1\approx p_c-\delta p,\qquad
u_\alpha=(u_\alpha)_c+\O\left((\delta p)^2\right)\quad(\alpha=2,\dots,M).
\label{juc}
\eeq
We thus have $M-Nu_1\approx N\,\delta p$, hence the estimate
\beq
H_{ij}(t)\approx\frac{1}{p_c^iq_c^{j-i}N\,\delta p}
\approx\frac{1}{p_c^iq_c^{j-i}}\left(\frac{t_c}{2Np_cq_c(t_c-t)}\right)^{1/2}.
\label{jhsg}
\eeq
This inverse-square-root singularity can alternatively be obtained
by means of eq.~(\ref{tauf}), from the large-$K$ behavior of the coefficients
of the series~(\ref{jhdef}).

We have thus shown that the functions $G^\pm_{k}(t)$ and $G^\pm_{k,\ell}(t)$
which enter the linear equations~(\ref{G=GF}) are algebraic in $t$,
with degree $\bin{N}{M}$.
They possess branch points of the form~(\ref{jhsg}) at the critical point
($t=t_c)$.
Unfortunately, these properties are not sufficient
to evaluate the critical behavior of the functions $F^\pm_k(t)$ near $t_c$.
Indeed, it can be checked that the linear system~(\ref{G=GF})
becomes very singular at $t=t_c$:
the most singular part of the matrix $G^\pm_{k,\ell}(t)$
as $t\to t_c$ is a matrix of rank one.

It is worth noting that the small roots
of the polynomial equation~(\ref{jupoly})
also play an important role in recent works~\cite{DLA}
devoted to large deviations in exclusion models.

\subsection{Probability flow and algebraic trick}
\label{sec:algpro}

An algebraic treatment can also be applied to eq.~(\ref{clauderat}).
Consider a fixed $p>p_c$, so that $t<t_c$.
If in eq.~(\ref{clauderat}) we replace
$p$ by any of the small roots $u_\alpha$ of eq.~(\ref{jupoly}),
the r.h.s. vanishes.
Indeed this expression has been shown below eq.~(\ref{jdefz})
to vanish for $\abs{z(p)}<z(p_c)$,
and we have indeed $\abs{z(u_\alpha)}=\abs{z(p)}<z(p_c)$.
Hence
\beq
u_\alpha=\sum_{k=1}^Mu_\alpha^k(1-u_\alpha)^{n^\pm_k-k}F^\pm_k(t)
\qquad(\alpha=1,\dots,M).
\label{jlinrat}
\eeq
This linear system of $M$ equations determines the $F^\pm_k(t)$.
Eq.~(\ref{clauderat}) then leads to the survival probabilities $F^\pm$.

This algebraic trick is quite powerful for numerical purposes.
Indeed, for given $p$, it only involves
solving the polynomial equation~(\ref{jupoly})
and the linear system~(\ref{jlinrat}).
Figure~10 shows a plot of the survival probabilities $F^\pm$
against the slope $v$, for different values of $p$, indicated on the curves.
For given $p$, the algebraic trick has been applied numerically for all
rational slopes having a denominator $N\le40$.
The numerical values of $F^\pm$ are thus obtained.
For a rational slope $v$, $F^\p$ (respectively, $F^\m$)
can be read as the ordinate just after
(respectively, just before) the discontinuity of the curve $F(v)$.

\vskip 8.5cm{\hskip 0.8cm}
\includegraphics{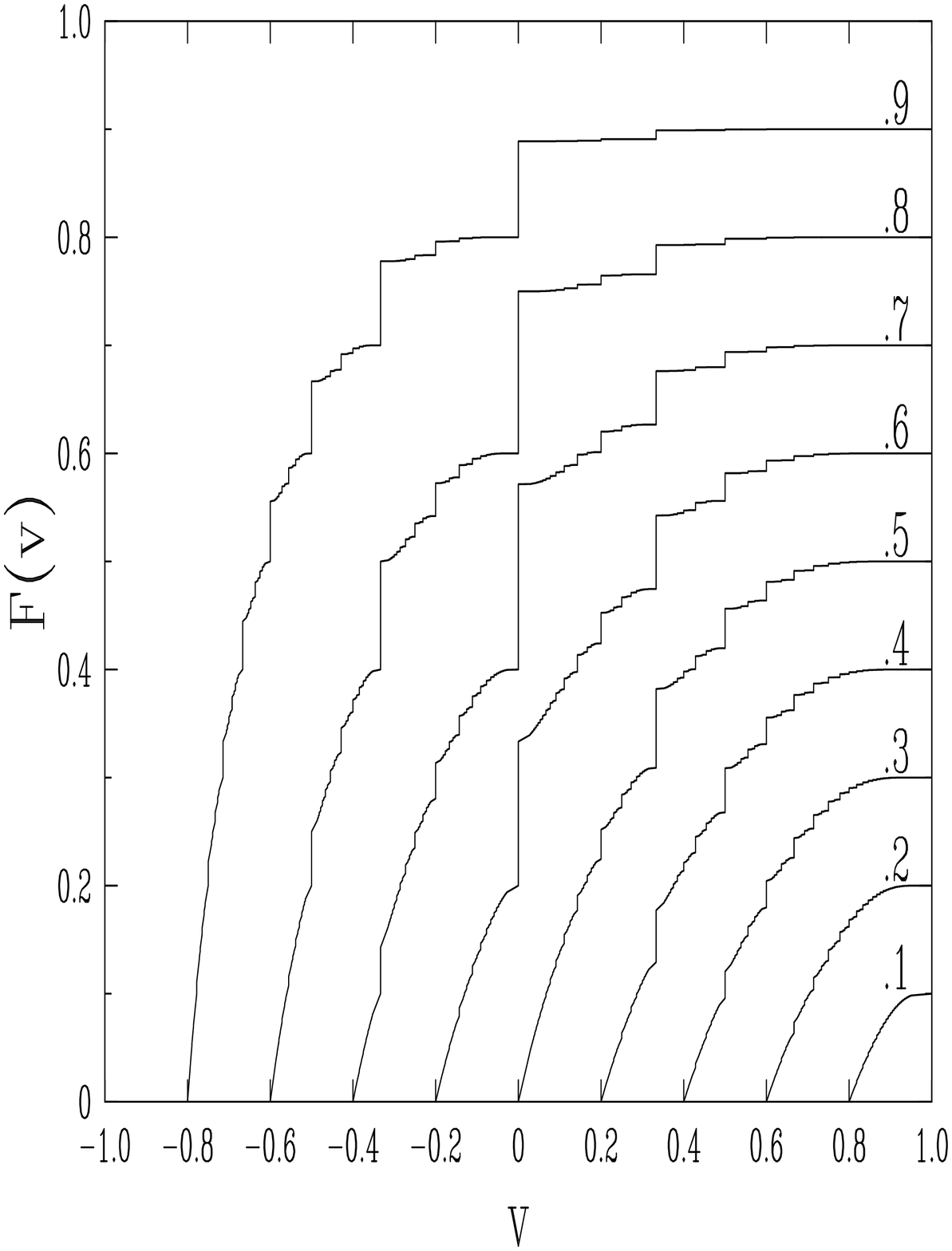}

\capt{10}{J2}{Plot of the survival probabilities $F^\pm$ against $v$,
for several values of $p$, indicated on the curves.
The apparent $v\to1$ limit of $F(v)$ is $F^-(1)=p$~[see eq.~(\ref{jsimple})],
whereas the jump at $v=1$ to $F^+(1)=1$ is not visible.}

The algebraic trick also demonstrates that the $F^\pm_k$
are generically algebraic functions of degree $\deg_t(M,N)$ in $t$,
while $F^\pm$
are generically algebraic functions of degree $\deg_p(M,N)$ in $p$, with
\beq
\deg_t(M,N)=\bin{N}{M},\qquad\deg_p(M,N)=\bin{N-1}{M}.
\label{jdegre}
\eeq
Elimination theory can be used to write down explicitly
the algebraic relations between $p$ or $t$ and the $F^\pm_k$ or
$F^\pm$.
Various examples will be given in sec.~\ref{sec:further}.

If we are interested only in $F^\pm$, there is an alternative
route\footnote{This approach was suggested to us by V.~Lafforgue.}.
Instead of computing the $F^\pm_k$, we can solve a transposed linear system,
whose meaning is the following.
Suppose that $p$ is replaced by one of the small roots $u_\alpha$
to count the weight of the walks.
Then the crossing edges that end on the wall get the correct weights,
while the other ones are slightly wrong,
because $n^\pm_k/k$ is only approximately equal to $N/M$.
By taking a linear combination of the weights corresponding to the $u_\alpha$,
we can give the right weight to any crossing edge.
Indeed, given the $u_\alpha$, we can determine numbers $D^\pm_\alpha$
such that we have, for any $k\ge1$
\beq
\sum_{\alpha=1}^M
D^\pm_\alpha u_\alpha^k(1-u_\alpha)^{n^\pm_k-k}=p^kq^{n^\pm_k-k}.
\label{jdefd}
\eeq
This is possible because eq.~(\ref{jdefd}) has to be imposed
only for $k=1,\dots,M$, as a consequence of eq.~(\ref{jnkper}).
We can substitute eq.~(\ref{jdefd}) into eq.~(\ref{claude}),
and use again the fact that the r.h.s. of eq.~(\ref{claude})
vanishes when we substitute $u_\alpha$ for $p$, obtaining thus
\beq
F^\pm=p-\sum_{\alpha=1}^MD^\pm_\alpha u_\alpha.
\eeq

We close up this section by investigating the relationship
between $F^\p$ and $F^\m$.
Using eq.~(\ref{jnklim}), we can recast eq.~(\ref{clauderat}) as
\beqa
F^\p&=&p(1+tF^\p_M)-\sum_{k=1}^{M-1}p^kq^{n^\pm_k-k}F^\p_k-tF^\p_M,\nonumber\\
F^\m&=&p-\sum_{k=1}^{M-1}p^kq^{n^\pm_k-k}F^\m_k-tF^\m_M.
\eeqa
Since the above equations determine the $F^\pm_k$,
a comparison between both equations yields
\beq
F^\m=\frac{F^\p}{1+tF^\p_M},\quad
F^\m_k=\frac{F^\p_k}{1+tF^\p_M},\qquad(k=1,\dots,M),
\label{jfmfp}
\eeq
or equivalently
\beq
F^\p=\frac{F^\m}{1-tF^\m_M},\quad
F^\p_k=\frac{F^\m_k}{1-tF^\m_M},\qquad(k=1,\dots,M).
\label{jfpfm}
\eeq
Finally, eqs.~(\ref{jq}) and~(\ref{jpi}) respectively become
\beq
Q(v)=tF^\m_M
\label{jqr}
\eeq
and
\beq
\Pi(v)=\frac{tF^\p_MF^\p}{1+tF^\p_M}=\frac{tF^\m_MF^\m}{1-tF^\p_M}
=tF^\m_MF^\p=tF^\p_MF^\m.
\label{jpir}
\eeq

\subsection{Duality}
\label{sec:algtrickdual}

We now turn to an investigation of the relationship
between the survival probability
associated with the rational slopes $v$ and~$-v$,
by exploring the duality approach of sec.~\ref{sec:du} in the rational case.
We notice that substituting $v$ for $-v$ amounts to exchanging
the roles of the integers $M$ and $\w M$ in the definitions~(\ref{jdefmn}).

We start with the following combinatorial observation.
Let $(t=n,x=n-2k)$ be a point with integer coordinates on the wall.
The number of finite walks starting at the origin, remaining on
the same side of the wall and ending at $(t,x)$
is the same for left walks and for right walks.
There is indeed an obvious one to one correspondence between both sets
of walks, depicted in Figure~11.
The second walk (contributing to right walks) is obtained from the first one
(contributing to left walks) by a rotation of angle $\pi$
around the mid-point $(t/2,x/2)$.

$$\centerline{\epsfbox{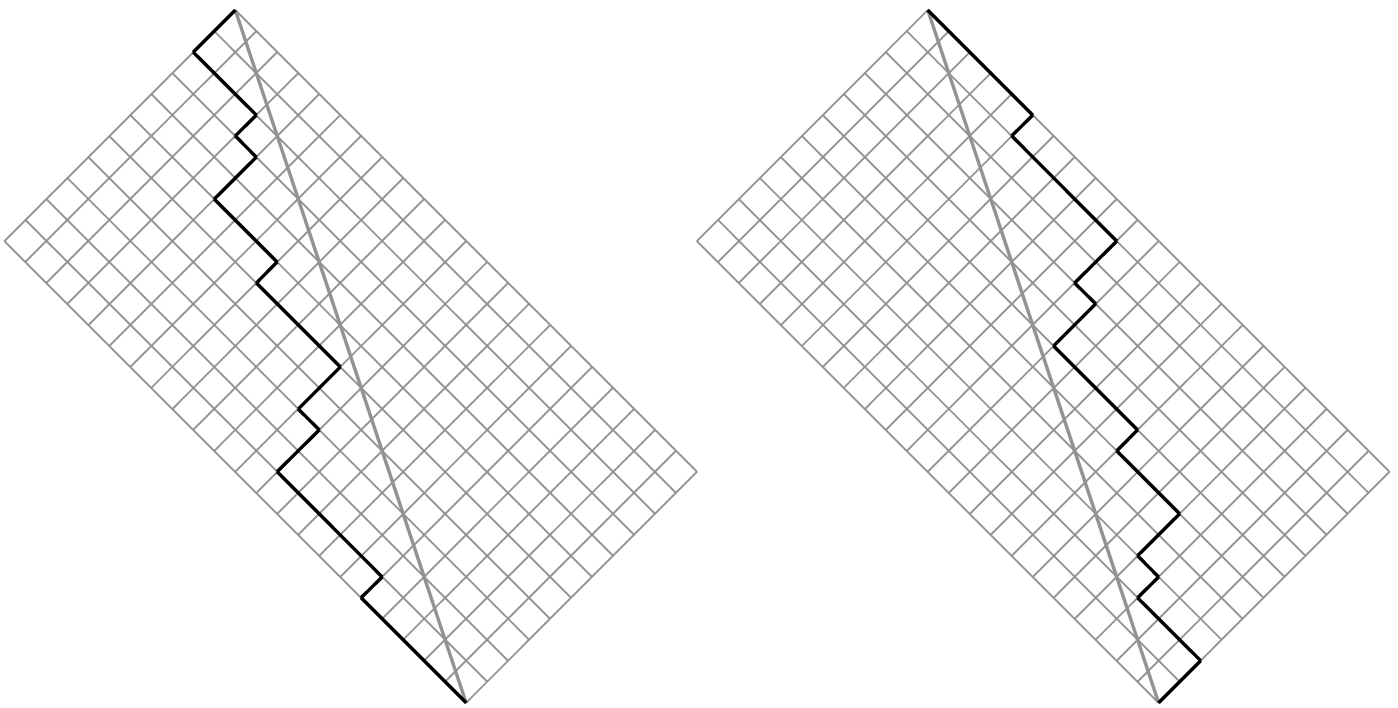}}$$
\capt{11}{M9}{Combinatorial proof of the identity~(\ref{jfid}).}

This observation implies the following identity between generating functions
\beq
F^\pm_M(t,v)=F^\pm_{\w M}(t,-v).
\label{jfid}
\eeq
In particular, the probability $Q$ that a walk crosses the wall and does
so for the first time at a point with integer coordinates,
as given by eq.~(\ref{jqr}), is the same for right and left walks.
We shall not use this information for a while,
but rather derive many other identities, including eq.~(\ref{jfid}),
by employing only algebraic means.

Following the same strategy as in the irrational case, we start from
eqs.~(\ref{jidp}),~(\ref{jidpp}), which now read
\beqa
&&p-\sum_{k=1}^Mp^kq^{n^\pm_k-k}F^\pm_k(t)=0
\qquad(0\le p\le p_c),
\label{jpr1}\\
&&p-\sum_{k=1}^{\w M}p^kq^{\w n^\pm_k-k}\w F^\pm_k(\w t)=0
\qquad(0\le p\le q_c),
\label{jpr2}
\eeqa
with $\w t=p^{\w M}q^M$.
If we now substitute $q=1-p$ for $p$ in eq.~(\ref{jpr2})
(this turns $\w t$ into $t$), and then multiply it
by eq.~(\ref{jpr1}), we get
\beq
\left(1-q-\sum_{k=1}^M p^kq^{n^\pm_k-k}F^\pm_k\right)
\left(q-\sum_{k=1}^{\w M}q^kp^{\w n^\pm_k-k}\w F^\pm_k\right)=0
\qquad(0\le p\le1).
\label{algtrickdual}
\eeq

We again use the canonical walk, but now only a finite piece of it.
For definiteness, we concentrate our attention on the $F^\p_k$,
for both slopes $v$ and~$-v$.
We first observe that $n^\p_M=\w n^\p_{\w M}=N+1$, so that, if we define
\beq
F^\p_0=1+tF^\p_M,\qquad\w F^\p_0=1+t\w F^\p_{\w M},
\eeq
eq.~(\ref{algtrickdual}) becomes
\beq
\left(1-qF^\p_0-\sum_{k=1}^{M-1}p^kq^{n^\p_k-k}F^\p_k\right)
\left(1-p\w F^\p_0-\sum_{k=1}^{\w M-1}q^kp^{\w n^\p_k-k}\w F^\p_k\right)=0
\qquad(0\le p\le1).
\label{jalgtrickdual}
\eeq

We now change notation for the unknowns inside the sums,
just as in the irrational case,
labeling the $F^\p$ functions by the value of $n_k$ or $\w n_k$,
rather than with $k$.
We employ the notations~(\ref{jkn})--(\ref{jdefx}),
albeit with the sets $I^\pm$ now being a partition of the integers
$\ell=2,\dots,N-1$.
Eq.~(\ref{jalgtrickdual}) reads
\beq
\left(1-qF^\p_0-\sum_{\ell\in I^+}X_\ell\oF^\p_\ell\right)
\left(1-p\w F^\p_0-\sum_{\ell\in I^-}X_\ell\oF^\p_\ell\right)=0
\qquad(0\le p\le1).
\eeq

The strategy for expanding this product is the following.
We substitute $t$ for $p^M(1-p)^{N-M}$ as often as we can.
For fixed $t$, we thus end up with a polynomial in $p$ of degree less than $N$.
Such a polynomial has to vanish identically, because it has $N$ roots.
Indeed, the first factor in eq.~(\ref{jalgtrickdual})
vanishes when $p$ is equal to any of the $M$ small roots $u_\alpha$
of the polynomial equation~(\ref{jupoly}),
while the second factor vanishes on the other $\w M$ roots.

Now consider $\ell\in I^+$ and $\ell'\in I^-$.
If $\ell+\ell'<N$, then $X_\ell X_{\ell'}=X_{\ell+\ell'}$.
If $\ell+\ell'=N$, then $X_\ell X_{\ell'}=t$.
If $\ell+\ell'>N+1$, then $X_\ell X_{\ell'}=tX_{\ell+\ell'-N}$.
The equality $\ell+\ell'=N+1$ never occurs.
We also have $pX_\ell=X_{\ell+1}$ if $\ell+1<N$
and $pX_{N-1}=t$ if $N-1\in I^+$, and symmetrically $qX_{\ell'}=X_{\ell'+1}$
if $\ell'+1<N$ and $q_{N-1}=t$ if $N-1\in I^-$.
Finally, $pq=p(1-p)=X_2$.
All these properties can be checked by combining the
inequalities defining $k_\ell$ and $k_{\ell'}$ [see eq.~(\ref{jkn})].
All the $X_\ell$'s have different degrees ranging from 2 to $N-1$, so that,
together with any two of the three polynomials $1$, $p$ and $1-p$,
they form a basis of the polynomials of degree less than $N$.
We concentrate for a while on terms of degree less than two in $p$.
We are left with the sum of a function of $t$ and of $-(1-p)\w F^\p_0-pF^\p_0$,
which has to vanish identically.
This implies in particular $\w F^\p_0=F^\p_0$,
which is just the identity~(\ref{jfid}),
that we have proved in the beginning of this section by combinatorial means.
We set
\beq
\oF^\p_1=\w F^\p_0=F^\p_0,
\eeq
and $I_0^+=I^+\cup\{1\}$ and $I_0^-=I^-\cup\{1\}$.
Just as in the irrational case, we meet many simplifications,
and we are left with
\beq
1-\oF^\p_1
-\sum_{\ell=2}^{N-1}\oF^\p_\ell X_\ell+\sum_{\ell\in I_0^+,\ \ell'\in I_0^-}
\oF^\p_\ell\oF^\p_{\ell'}t^{\rho(\ell,\ell')}
X_{\ell+\ell'-N\rho(\ell,\ell')}=0,
\eeq
with the convention $X_0=1$, and with
\beq
\rho(\ell,\ell')=\left\{\matrix{
0&\hbox{if}\,\,\ell+\ell'<N,\hfill\cr
1&\hbox{if}\,\,\ell+\ell'\ge N.\hfill\cr
}\right.%}
\eeq
A term-by-term identification then leads to
\beq
\label{dualF1}
\oF^\p_1=1
+t\sum\dbli{\ell\in I_0^-,\ \ell'\in
I_0^+}{\ell+\ell'=N}\oF^\p_\ell\oF^\p_{\ell'}
\eeq
and
\beq
\label{dualFl}
\oF^\p_\ell
=\sum\dbli{\ell'\in I_0^-,\ \ell''\in
I_0^+}{\ell'+\ell''=\ell\,\mathrm{mod}\,N}
\oF^\p_{\ell'}\oF^\p_{\ell''}t^{\rho(\ell',\ell'')}
\qquad(\ell=2,\dots,N-1).
\eeq

Eqs.~(\ref{dualF1}) and~(\ref{dualFl}) provide $N-1$ equations
for the $N-1$ unknowns $\oF^\p_\ell$.
When $t$ goes to 0, eq.~(\ref{dualF1}) becomes trivial,
while the other ones become identical to the recursion relation~(\ref{rec3}).
As any irrational slope $v$ can be approximated
by a rational one that leads to the same coefficients $\oA_\ell$
up to an arbitrary given $\ell$,
we thus obtain a proof of the third recursion relation~(\ref{rec3})
by algebraic means, without any recourse to analysis.

\section{Critical behavior}
\label{sec:algcrit}

This section is devoted to the analysis of the `critical behavior'
as $p\to p_c$ of the survival probabilities $F^\pm$
and of related quantities in the convergent regime $(p>p_c)$,
for a fixed slope $v$, either rational or not.
This investigation also yields quantitative predictions
concerning the amplitudes $C^\pm(v)$ in the marginal regime $(p=p_c)$
and the prefactors $b^\pm(np_c)$ in the large-deviation regime $(p<p_c)$.

\subsection{Convergent regime}

We consider first the convergent regime, for a rational slope.
The behavior~(\ref{jasya}) of the integers $A^\pm_k$ implies that
the generating series $F^\pm_k(t)$
have square-root singularities at $t=t_c$,
as shown by eq.~(\ref{tauf}).
We again employ the notations~(\ref{jcrit}),
and we define the positive amplitudes $\Phi_k^\pm$ such that
\beq
F^\pm_k(t)\approx(F^\pm_k)_c-\Phi_k^\pm\abs{\delta p}
\qquad(k=1,\dots,M,\,\,\delta p\to 0),
\label{jfkc}
\eeq
with $(F^\pm_k)_c=F^\pm_k(t_c)$.
Similarly, the survival probabilities $F^\pm$ are expected to vanish
linearly as $p\to p_c+0$, with amplitudes $\Phi^\pm$:
\beq
F^\pm\approx\Phi^\pm\,\delta p\qquad(\delta p\to+0).
\eeq

The critical amplitudes $\Phi^\pm$ and $\Phi_k^\pm$
can be determined by means of the algebraic trick as follows.
By differentiating eq.~(\ref{clauderat}), we obtain
\beq
\frac{\d F^\pm}{\d p}=1-\sum_{k=1}^Mp^kq^{n^\pm_k-k}
\left\{\left(\frac{k}{p}-\frac{n^\pm_k-k}q\right)F^\pm_k
+\frac{\d F^\pm_k}{\d p}\right\}.
\label{jcrd}
\eeq
Now, by inserting the limits
\beq
\matrix{
\frad{\d F^\pm}{\d p}\to\Phi^\pm,\hfill&
\frad{\d F^\pm_k}{\d p}\to-\Phi^\pm_k\hfill&
(p\to p_c+0),\hfill\cr
\frad{\d F^\pm}{\d p}\to0,\hfill&
\frad{\d F^\pm_k}{\d p}\to\Phi^\pm_k\hfill&
(p\to p_c-0),\hfill\cr
}
\eeq
into eq.~(\ref{jcrd}), and using eq.~(\ref{jnpm}),
we obtain two different expressions for the amplitudes $\Phi^\pm$:
\beq
\Phi^\pm=2\left(1+\sum_{k=1}^Mp_c^kq_c^{n^\pm_k-k-1}
\left(1-\Frac^\mp(k/p_c)\right)(F^\pm_k)_c\right)
=2\sum_{k=1}^Mp_c^kq_c^{n^\pm_k-k}\Phi_k^\pm.
\label{jphi2}
\eeq

The middle side of eq.~(\ref{jphi2}) gives $\Phi^\pm$ in terms
of the $(F^\pm_k)_c$, which can be obtained by the algebraic trick at $p=p_c$.
The rightmost side of eq.~(\ref{jphi2}) then allows to determine
the $\Phi_k^\pm$ by means of a second use of the algebraic trick.
Indeed the behavior~(\ref{juc}) of the roots $u_\alpha$ implies
$\d u_\alpha/\d p\to-\delta_{\alpha,1}$ in the $p\to p_c+0$ limit.
As a consequence, the $\Phi_k^\pm$ can be determined from the following
linear system of $M$ equations
\beq
\frac{\delta_{\alpha,1}}{2}\Phi^\pm
=\sum_{k=1}^M(u_\alpha)_c^k(1-(u_\alpha)_c)^{n^\pm_k-k}\Phi_k^\pm
\qquad(\alpha=1,\dots,M).
\label{jlinp}
\eeq

It turns out that the amplitudes $\Phi^\pm$ and $\Phi_k^\pm$ govern the
behavior of various quantities for a rational slope.
First, eq.~(\ref{tauf}) yields the asymptotic behavior
of the integers $A^\pm_k$ for large $k$.
We thus recover eq.~(\ref{jasya}),
and we obtain expressions for the amplitudes $B^\pm_k$,
which asymptotically only depend on $k$ modulo $M$:
\beq
B^\pm_{KM+k}\approx
Np_cq_c\,(t_c)^{k/M}\,\Phi_k^\pm\qquad(k=1,\dots,M,\,\,K\gg1).
\label{jasb}
\eeq
This result can be recast as follows.
Since $p_c=M/N$ is rational, the fractional part
$\Frac^\pm(k/p_c)=\Frac^\pm(kN/M)$
takes $M$ rational values with denominator $M$,
and only depends on $k$ modulo $M$.
Conversely, any function of $k$ modulo $M$
can be considered as a function of $\Frac^\pm(k/p_c)$.
Hence eq.~(\ref{jasb}) is equivalent to
\beq
B^\pm_k\approx B^\pm(k/p_c),
\label{jbper}
\eeq
i.e.,
\beq
A^\pm_k\approx B^\pm(k/p_c)\,\frac{p_c}{(2\pi q_c\,k^3)^{1/2}}
\,\left(p_c\,q_c^{q_c/p_c}\right)^{-k}\qquad(k\gg1),
\label{jasyaper}
\eeq
where $B^\pm(x)$ are periodic functions, with unit period.
Since the functional form of eq.~(\ref{jasyaper}) holds
independently of the slope $v$, provided $v$ is rational,
we make the reasonable hypothesis that the law~(\ref{jasyaper})
also holds when $v$ is irrational,
albeit with one single periodic function $B(x)$.

The bilateral sequence of integers $\oA^\pm_n$,
introduced in eq.~(\ref{jdefan}), can be argued to exhibit
an asymptotic behavior analogous to eq.~(\ref{jasyaper}):
\beq
\oA^\pm_n\approx\beta^\pm(\theta_n)\,\frac{1}{(2\pi p_cq_c\,n^3)^{1/2}}
\,\left(p_c^{p_c}q_c^{q_c}\right)^{-n}\qquad(n\gg1),
\label{jasyadu}
\eeq
where the arguments $\theta_n$ of the periodic functions $\beta^\pm(\theta)$
have been defined in eq.~(\ref{jkn}).
The estimate~(\ref{jasyadu}), with its periodic functions $\beta^\pm(\theta)$,
involves $p_c$ and $q_c$ in a symmetric fashion,
contrary to eq.~(\ref{jasyaper}).
Furthermore it encodes simultaneously the asymptotic behavior
of the sequences $A^\pm_k$ associated with both slopes $\pm v$.
These are advantages of the duality approach.
Eq.~(\ref{jthplus}) also implies the following relationship
\beq
B^\pm(x)=\left(p_c\,q_c^{q_c/p_c}\right)^{-\theta}\beta^\pm(\theta),
\qquad\theta=p_c(1-x)\qquad(0\le\theta\le p_c),
\eeq
between the periodic functions of the estimates~(\ref{jasyaper})
and~(\ref{jasyadu}), with corresponding arguments.
Figures~12 and~13 respectively show plots of the periodic functions
$B(x)$ and $\beta(\theta)$,
again for the golden slope given in eq.~(\ref{jgold}).

\subsection{Marginal and large-deviation regimes}

Quantities pertaining to the marginal regime $(p=p_c)$
and to the large-deviation regime $(p<p_c)$
can also be characterized in terms of the
critical amplitudes $\Phi^\pm$ and $\Phi_k^\pm$, for a rational slope.

In these two regimes, the limit survival probabilities $F^\pm$ vanish,
so that the probability flow equation~(\ref{jcn1}) becomes
\beq
F^\pm(n,v)=\sum_{k\ge\Int^\pm(np_c)+1}A^\pm_k p^kq^{n^\pm_k-k}.
\label{jcn2}
\eeq

\vskip 8.5cm{\hskip 0.8cm}
\includegraphics{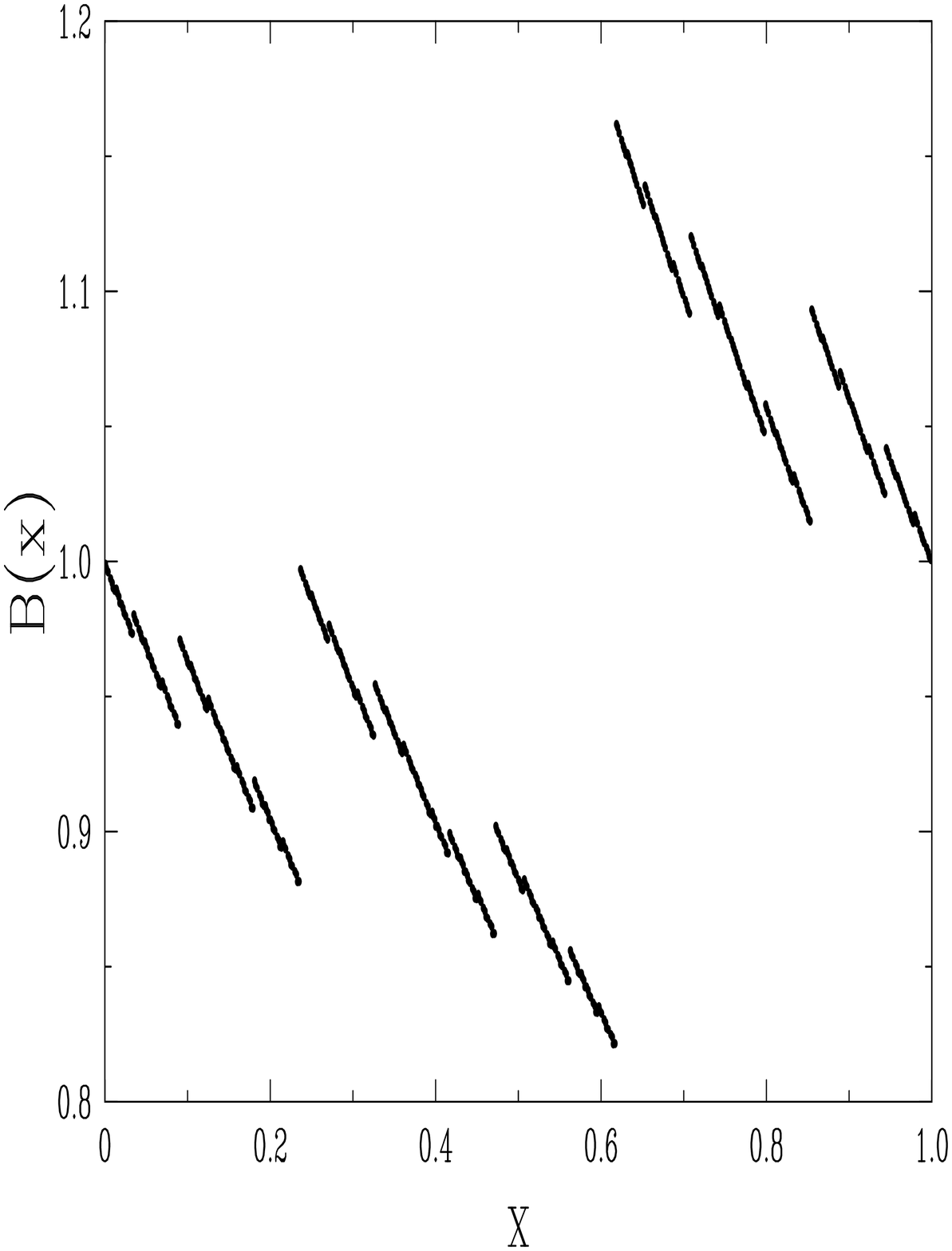}

\capt{12}{J7}{Plot of the periodic amplitude $B(x)$,
for the golden slope.}

\vskip 8.5cm{\hskip 0.8cm}
\includegraphics{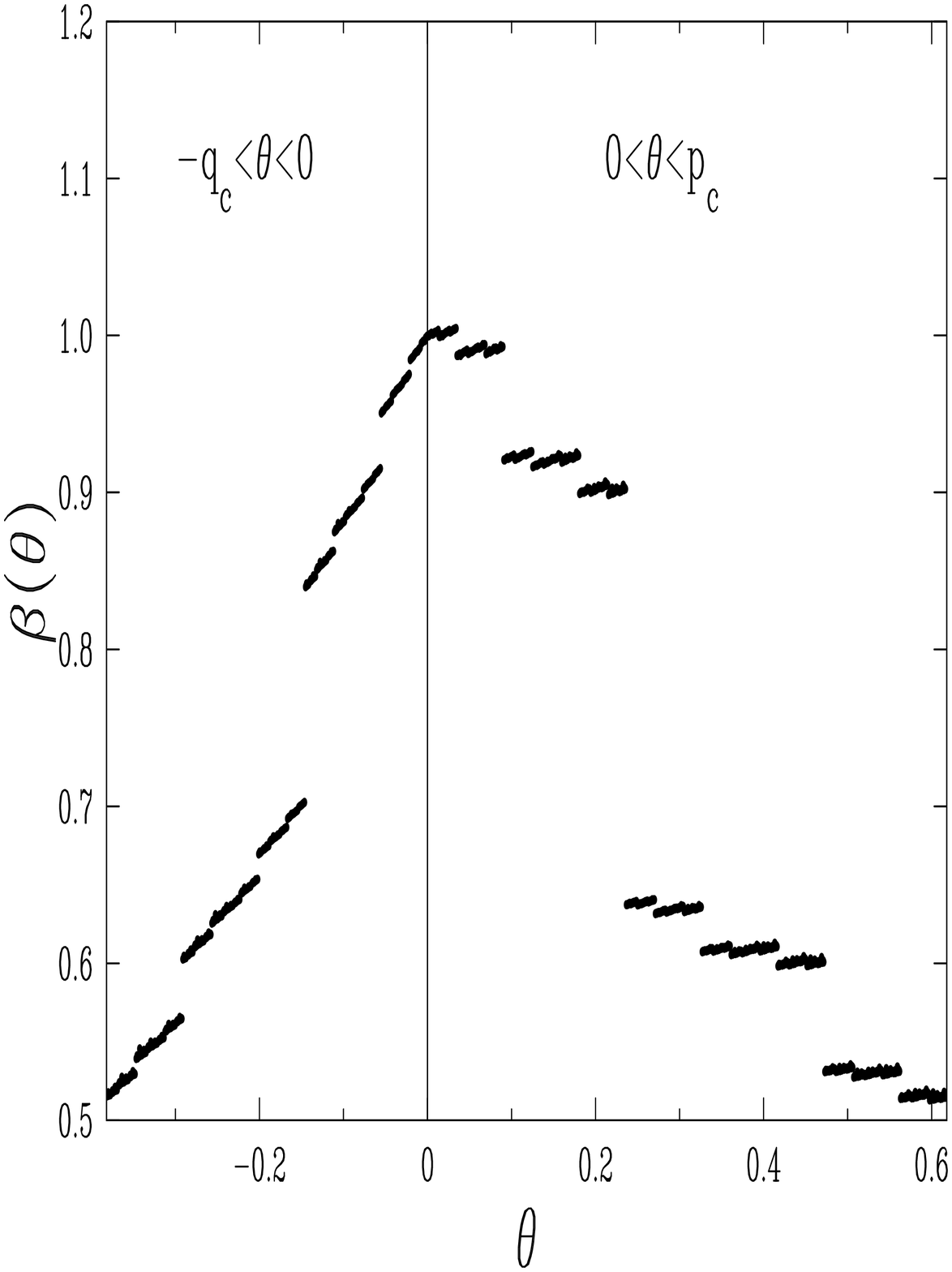}

\capt{13}{J8}{Plot of the periodic amplitude $\beta(\theta)$,
for the golden slope.}

In the marginal regime $(p=p_c)$,
the behavior of $F^\pm(n,v)$ for large $n$ can be derived
by inserting the estimates~(\ref{jasya}) and~(\ref{jasb})
into eq.~(\ref{jcn2}).
The terms in the r.h.s. of that equation decay as the power law $k^{-3/2}$,
so that the result is rather insensitive to the details of the sequences
$A^\pm_k$.
Indeed the amplitudes $\Phi^\pm_k$ only appear through
the combinations which enter the rightmost side of eq.~(\ref{jphi2}).
We thus recover the general result~(\ref{jfc}), with an amplitude
\beq
C^\pm(v)=\left(\frac{p_cq_c}{2}\right)^{1/2}\Phi^\pm
=\left(\frac{1-v^2}{8}\right)^{1/2}\Phi^\pm,
\eeq
in agreement with eq.~(\ref{jfpc}).

\vskip 8.5cm{\hskip 0.8cm}
\includegraphics{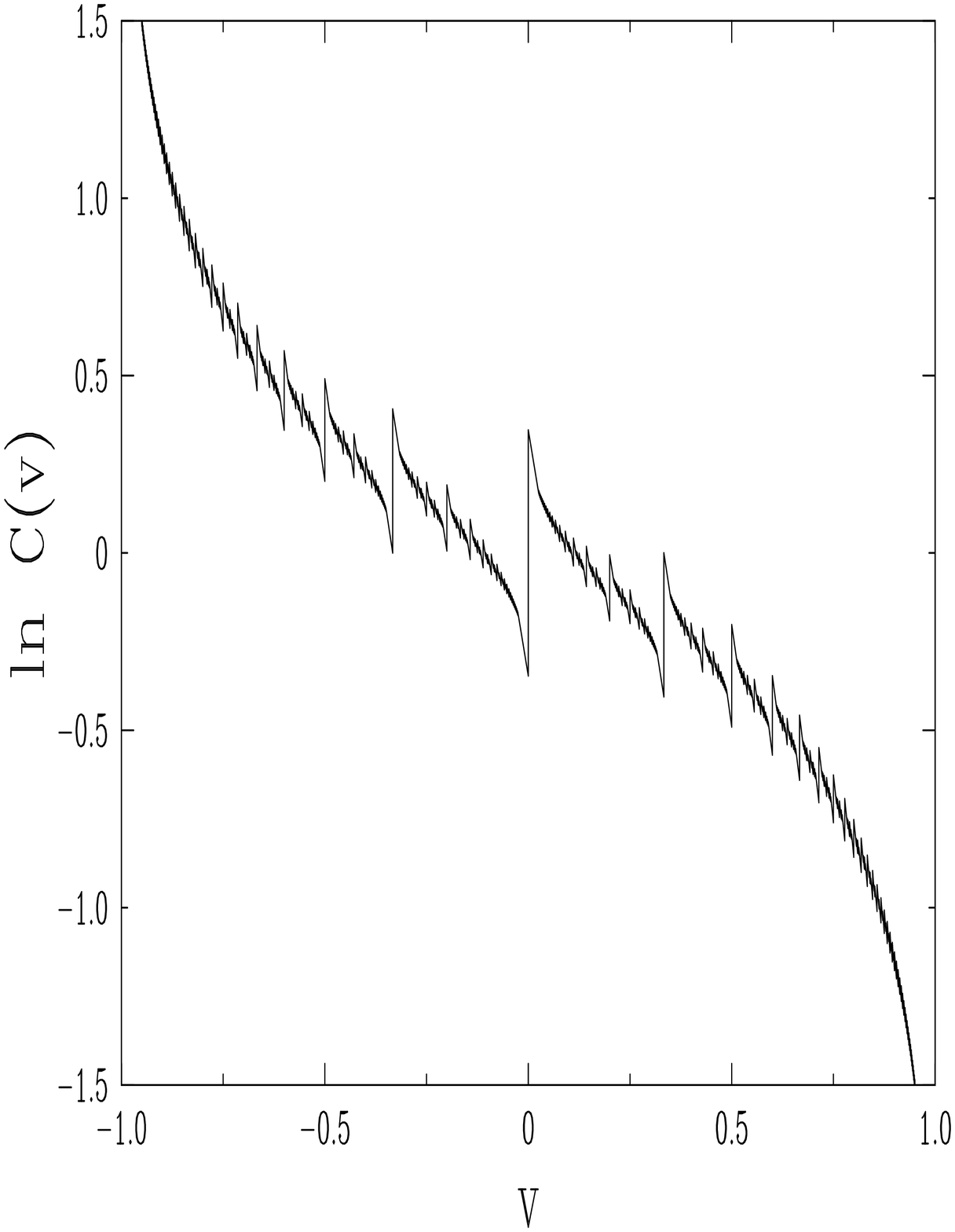}

\capt{14}{J5}{Logarithmic plot of the critical amplitude $C(v)$ against $v$.}

The algebraic trick at the critical point thus allows
a numerical determination of the amplitude $C(v)$ of the law~(\ref{jfc}).
Figure~14 shows a logarithmic plot of this amplitude,
obtained from data at rational slopes, along the lines of Figure~10.
The curve is centrally symmetric,
as a consequence of the duality symmetry~(\ref{jduc}).
Figure~15 shows a similar, albeit more appealing, plot of the combination
\beq
y^\pm(v)=\ln C^\pm(v)+\frac{1}{2}\ln\frac{q_c}{p_c}
=\ln C^\pm(v)+\frac{1}{2}\ln\frac{1+v}{1-v}.
\label{jy}
\eeq
This quantity obeys the following bounds,
shown as dashed lines in Figure~15:
\beqa
-\frad{\ln 2}{2}\le&y^\pm(v)&\le\frad{\ln 2}{2}-\ln(1-v)\qquad
(-1\le v\le0),\nonumber\\
\ln(1+v)-\frad{\ln 2}{2}\le&y^\pm(v)&\le\frad{\ln 2}{2}
{\hskip 3cm}(0\le v\le1).
\label{jyb}
\eeqa
The results~(\ref{j1crit}) and~(\ref{jd1crit}),
to be derived later, show that the bounds~(\ref{jyb})
are saturated by the slopes with either $M=1$ or $\w M=1$.
These bounds lead to the limit values
\beq
y(\pm1)=\pm\frac{\ln2}{2}.
\label{jylim}
\eeq

\null\vskip 8.5cm{\hskip 0.8cm}
\includegraphics{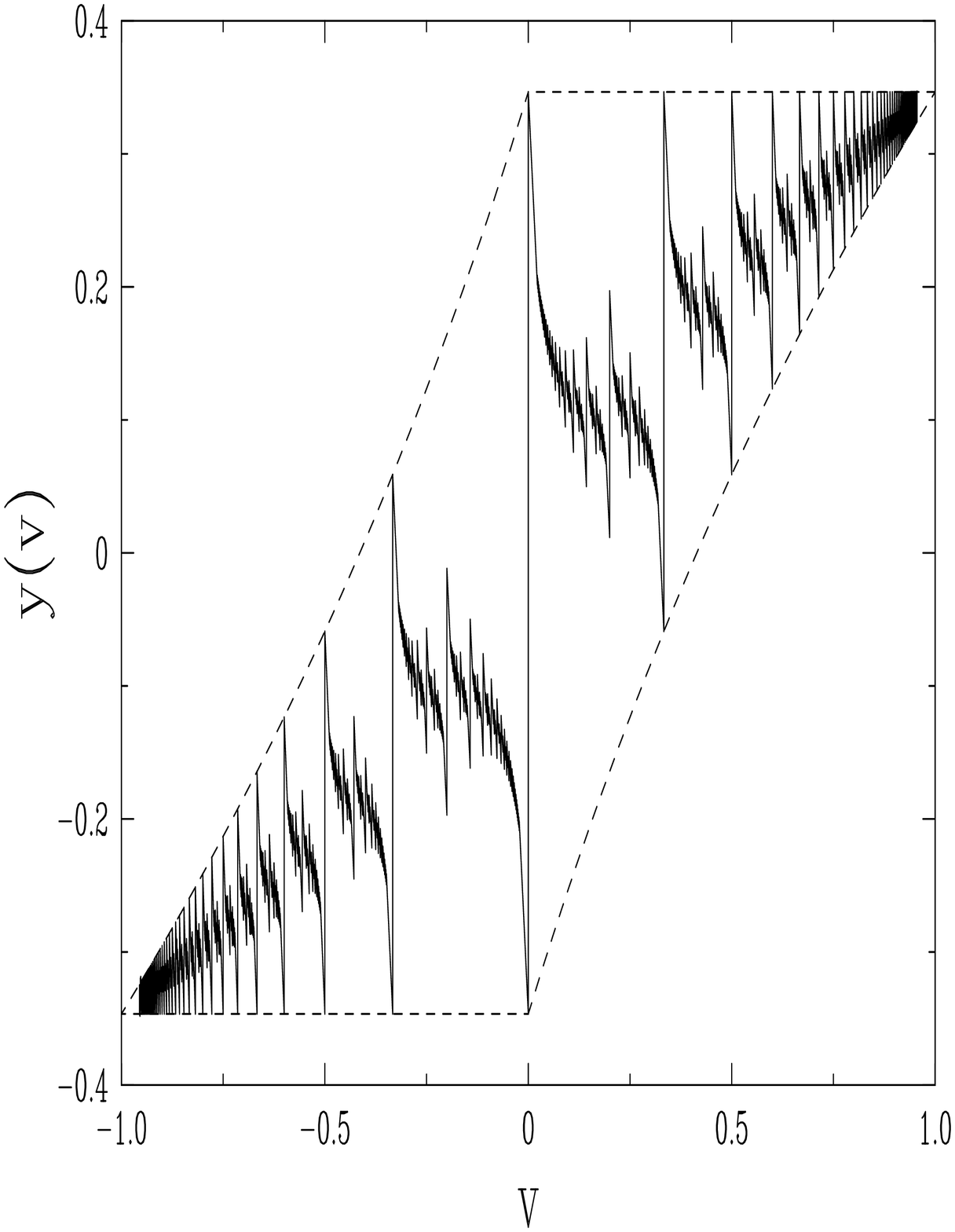}

\capt{15}{J6}{Plot of $y(v)$, defined in eq.~(\ref{jy}), against $v$.}

The investigation of the large-deviation regime $(p<p_c)$ is more involved.
The sum in the r.h.s. of eq.~(\ref{jcn2}) is now exponentially convergent,
so that it is dominated by the fine structure of the $A^\pm_k$
near the lower bound, i.e., for $k\approx np_c$.
We recover after some manipulations the general result~(\ref{jasyf}),
with amplitudes which only depend on $n$ modulo $N$:
\beq
b^\pm_{KN+n}\approx\frac{Nq_c}{p_c^{1/2}(1-\e^{-NS(v)})}
\sum_{k=1}^M q^{1-\Frac^\mp(k/p_c)}
\,\e^{[(n-k/p_c)-N\Theta(n-k/p_c)]S(v)}\,(t_c)^{k/M}\Phi^\pm_k,
\eeq
for $n=1,\dots,N$ and $K\gg1$, where $\Theta$ is the Heaviside step function,
and $S(v)$ the entropy function of eq.~(\ref{js}).
This result agrees with the existence of a periodic prefactor
to the law~(\ref{jasyf}),
since any function of $n$ modulo $N$ can be viewed as a function of
$\Frac^\pm(np_c)=\Frac^\pm(nM/N)$.

\section{Further results in specific cases}
\label{sec:further}

\subsection{The limit case $v=1$, i.e., $p_c=0$}
\label{sec:vun}

In this limit situation,
quantities with the superscript $\p$ are equal to the same quantities
for the whole unconstrained set of random walks.
Quantities with the superscript $\m$ are only constrained
by the condition that the first step is to the left $(\eps_1=-1)$.

The general formalism of this paper agrees with these considerations.
Indeed, any value of $p>0$ corresponds to the convergent regime.
Eq.~(\ref{jpsum}) implies that the one-time distribution functions read
\beq
P^\p(n,v)=1,\quad P^\m(n,v)=1-q^n\qquad(n\ge1),
\eeq
in agreement with the value $S=-\ln q$ of the entropy function
[see eq.~(\ref{js})].
We have therefore
\beq
f^\p(z,v)=\frac{1}{1-z},\qquad f^\m(z,v)=\frac{1-qz}{1-z},
\eeq
so that
\beq
F^\p(n,v)=1,\quad F^\m(n,v)=p\qquad(n\ge1).
\eeq
The integers $\N^\pm_{n,k}$ are also remarkably simple:
\beq
\N^\p_{n,k}=\bin{n}{k},\quad\N^\m_{n,k}=\bin{n-1}{k-1}\qquad(n\ge1,\ k\ge1).
\eeq
The limit survival probabilities read therefore
\beq
F^\p=1,\qquad F^\m=p,\qquad\Pi=q,
\label{jsimple}
\eeq
which can be recovered by elementary considerations:
we have $\Pi=\prob\big\{V=1\big\}=\prob\big\{\eps_1=1\big\}=q$
and $F^\p=1$, hence $F^\m=F^\p-\Pi=1-q=p$.

\subsection{The limit case $v=-1$, i.e., $p_c=1$}
\label{sec:vmun}

This situation is the dual of the previous one.
Quantities with the superscript $\m$ vanish identically,
while quantities with the superscript $\p$
have the single contribution of the walk
with all steps to the left $(\eps_n=-1$ for all $n)$.

Any value of $p<1$ now corresponds to the large-deviation regime.
Eq.~(\ref{jpsum}) leads to
\beq
P^\p(n,v)=p^n,\quad P^\m(n,v)=0\qquad(n\ge1),
\eeq
in agreement with the value $S=-\ln p$ of the entropy function.
We have therefore
\beq
f^\p(z,v)=\frac{1}{1-pz},\qquad f^\m(z,v)=0,
\eeq
so that the survival probabilities read
\beq
F^\p(n,v)=p^n,\quad F^\m(n,v)=0\qquad(n\ge0).
\eeq
The only non-zero integers $\N^\pm_{n,k}$ are
\beq
\N^\p_{n,n}=1\qquad(n\ge0).
\eeq

\subsection{Rational slopes with $M=1$, i.e., $v=1-2/N$ or $p_c=1/N$}
\label{sec:vemun}

The class of rational slopes with $M=1$
owes its simplicity to the fact that only one unknown function,
$F^\pm_1(t)$, is involved in the algebraic trick approach.

For $M=1$ and $N\ge2$, we have $v=1-2/N$, $p_c=1/N$, $q_c=(N-1)/N$,
$t=pq^{N-1}$, and $t_c=(N-1)^{N-1}/N^N$.
In the convergent regime $(p>p_c)$,
the polynomial equation~(\ref{jupoly}) has only one small root,
$u_1$, which is in the interval $0<u_1<p_c$.
Eq.~(\ref{jlinrat}) has the explicit solutions
\beq
F^\m_1=\frac{u_1}{t},\qquad F^\p_1=\frac{u_1}{t(1-u_1)}.
\label{j1f}
\eeq
We thus obtain
\beq
F^\m=p-u_1,\qquad F^\p=\frac{p-u_1}{1-u_1},
\qquad Q=u_1,\qquad\Pi=\frac{u_1(p-u_1)}{1-u_1}.
\label{j1sol}
\eeq
We notice that $u_1(t)$ and $F^\pm_1(t)$ are algebraic functions
of degree $N$ in $t$,
while $u_1(p)$ and $F^\pm(p)$ are algebraic functions
of degree $N-1$ in $p$, in agreement with eq.~(\ref{jdegre}).

The critical behavior of the quantities given in eqs.~(\ref{j1f}),
(\ref{j1sol}) is characterized by the amplitudes
\beqa
(F^\m_1)_c&=&\left(\frac{N}{N-1}\right)^{N-1},\qquad
(F^\p_1)_c=\left(\frac{N}{N-1}\right)^N,\nonumber\\
\Phi^\m_1&=&\frac{N^N}{(N-1)^{N-1}},\qquad
\Phi^\p_1=\frac{N^{N+2}}{(N-1)^{N+1}},\nonumber\\
\Phi^\m&=&2,\quad\Phi^\p=\frac{2N}{N-1},\qquad
C^\m=\frac{(2(N-1))^{1/2}}{N},\qquad
C^\p=\left(\frac{2}{N-1}\right)^{1/2}.
\label{j1crit}
\eeqa

The integers $A^\pm_k$ can be obtained in several ways.
The easiest route consists in evaluating them as contour integrals
of the generating series $F^\pm_1(t)$.
The expressions~(\ref{j1f}) yield
\beqa
A^\m_k&=&\oint\frac{\d t}{2\pi\i t^{k+1}}\,u_1
=\oint\frac{\d u_1}{2\pi\i u_1^k}\,\frac{1-Nu_1}{(1-u_1)^{Nk-k+1}},\nonumber\\
A^\p_k&=&\oint\frac{\d t}{2\pi\i t^{k+1}}\,\frac{u_1}{1-u_1}
=\oint\frac{\d u_1}{2\pi\i u_1^k}\,\frac{1-Nu_1}{(1-u_1)^{Nk-k+2}}.
\label{j1ak}
\eeqa
The integrals over $t$ are transformed to integrals over $u_1$
by means of the identity $t=u_1(1-u_1)^{N-1}$,
whereas the latter are linear combinations of the integrals over $u$
which enter eq.~(\ref{jintij}).
We thus obtain the integers $A^\pm_k$ as ratios of factorials:
\beq
A^\m_k=\frac{(Nk-2)!}{k!\,(Nk-k-1)!},\qquad
A^\p_k=\frac{(Nk)!}{k!\,(Nk-k+1)!}.
\label{jabin}
\eeq
These formulas can also be obtained combinatorially,
by means of the Raney principle~\cite{GKP}.
They agree with the observation, made below eq.~(\ref{bounds}),
that the $A^\m_k$ saturate the lower bound,
while the $A^\p_k$ saturate the upper bound.
The corresponding amplitudes $B^\pm_k$ read asymptotically
\beq
B^\m_k\approx q_c=\frac{N-1}{N},
\qquad B^\p_k\approx\frac{1}{q_c}=\frac{N}{N-1}.
\eeq

The situation of an immobile obstacle, i.e., of a vertical wall,
with slope $v=0$, corresponds to $M=\w M=1$ and $N=2$.
In this case, we have $p_c=1/2$ and $u_1=1-p$ for $p>1/2$,
so that the results~(\ref{j1f})--(\ref{j1sol}) can be further simplified to
\beqa
F^\m_1&=&\frac{1}{p},\qquad F^\p_1=\frac{1}{p^2},\qquad F^\m=2p-1,
\qquad F^\p=\frac{2p-1}{p},\nonumber\\
Q&=&1-p,\qquad\Pi=\frac{(2p-1)(1-p)}{p}.
\label{jverti}
\eeqa
The discontinuity $\Pi$ is maximal for $p=1/\sqrt2$, where it equals
\beq
\Pi\max=(\sqrt2-1)^2=0.171573.
\eeq
This value is the largest possible discontinuity met in this problem,
except the trivial one at $v=1$, namely $\Pi=q$ [see eq.~(\ref{jsimple})].

This example can be alternatively investigated ab initio.
Indeed, the integers $\N^\pm_{n,k}$ which obey eq.~(\ref{ntrat})
can be written as differences of binomial coefficients:
\beq
\N^\m_{n,k}=\bin{n-1}{k-1}-\bin{n-1}{k},\qquad
\N^\p_{n,k}=\bin{n}{k}-\bin{n}{k+1}.
\eeq
Eq.~(\ref{jan}) then yields,
in agreement with eq.~(\ref{jabin}) for $N=2$,
\beqa
A^\m_k&=&\N^\m_{2k-1,k}=\bin{2k-2}{k-1}-\bin{2k-2}{k}
=\frac{(2k-2)!}{k!\,(k-1)!}=\C_{k-1},\nonumber\\
A^\p_k&=&\N^\p_{2k,k}=\bin{2k}{k}-\bin{2k}{k+1}
=\frac{(2k)!}{k!\,(k+1)!}=\C_k,
\eeqa
where the $\C_k$ are the Catalan numbers~\cite{GKP}.

\subsection{Rational slopes with $\w M=1$, i.e., $v=-1+2/N$ or $p_c=1-1/N$}
\label{sec:vwemun}

The class of walls with $\w M=1$, i.e., a slope $v=2/N-1$,
gives a nice illustration of the duality approach,
exposed in sec.~\ref{sec:algtrickdual}.

In this situation, we have $M=N-1$, $p_c=1-1/N$, $q_c=1/N$,
$t=p^{N-1}q$, and $t_c=(N-1)^{N-1}/N^N$.
According to the general rules of sec.~\ref{sec:algtrickdual},
we introduce $N-1$ functions $\oF^\p_1,\dots,\oF^\p_{N-1}$.
The function $\oF^\p_1$ pertains to both $v$ and~$-v$,
while all the other ones only pertain to $v$.
We have $I_0^-=\{1\}$ and $I_0^+=\{1,\dots,N-1\}$,
so that the duality equations~(\ref{dualF1}), (\ref{dualFl}) respectively read
\beqa
\oF^\p_1&=&1+t\oF^\p_1\oF^\p_{N-1},\nonumber\\
\oF^\p_k&=&\oF^\p_1\oF^\p_{k-1}\quad\qquad(k=2,\dots,N-1).
\eeqa
So, we have $\oF^\p_k=\left(\oF^\p_1\right)^k$ for $k=1,\dots,N-1$, and
\beq
\oF^\p_1=1+t\left(\oF^\p_1\right)^N.
\label{jfpoly}
\eeq
The solution $\oF^\p_1(t)$ is the branch of eq.~(\ref{jfpoly})
which is regular at $t=0$.
A comparison of eq.~(\ref{jfpoly})
with the identity $t=p^{N-1}q$ shows that $\oF^\p_1=1/p$.
We thus have
\beq
\oF^\p_k=\frac{1}{p^k}\qquad(k=1,\dots,N-1).
\eeq
In order to obtain a complete combinatorial description of the problem,
we must recall the relation between the functions $\oF^\p_k$ and the functions
$F^\p_k$.
We have
\beq
n^\p_k=k+1\qquad(k=1,\dots,N-2),\qquad n^\p_{N-1}=N+1,
\label{jnkm}
\eeq
so that $\oF^\p_1=1+tF^\p_m=1+t\w F^\p_1$
and $\oF^\p_k=F^\p_{k-1}$ for $k=2,\dots,N-1$.
We thus obtain the remarkably simple result
\beq
F^\p_k=\frac{1}{p^{k+1}}\qquad(k=1,\dots,N-1).
\label{jd1k}
\eeq
Eqs.~(\ref{clauderat}) and~(\ref{jfmfp})--(\ref{jpir}) then yield
\beq
F^\m_k=\frac{1}{p^k}\qquad(k=1,\dots,N-1)
\label{jd1fk}
\eeq
and
\beq
F^\m=Np-(N-1),\quad F^\p=\frac{Np-(N-1)}{p},\quad
Q=q,\quad\Pi=\frac{q(Np-(N-1))}{p}.
\label{jd1gen}
\eeq

Eqs.~(\ref{jd1k})--(\ref{jd1gen}) show that the $F^\pm_k(t)$
are algebraic functions of degree $N$ in $t$, just as $p(t)$,
while $F^\pm(p)$ are rational functions, i.e., algebraic functions
of degree $1$ in $p$, in agreement with eq.~(\ref{jdegre}).
Finally, the results~(\ref{jverti}) can be recovered by setting $N=2$
in eqs.~(\ref{jd1k})--(\ref{jd1gen}).

The simplicity of the results~(\ref{jd1k})--(\ref{jd1gen}),
especially for $F^\m$, suggests that they can be obtained by elementary means.
This is indeed the
case\footnote{The following argument is due to V.~Lafforgue.}.
Fix $p>p_c=M/N$.
The probability to start to the right and to touch the wall is $q=1-p$,
because if the first step is to the right (probability $q$),
then the walk crosses the wall with probability one.
Let $R$ be the probability to start to the left and touch the wall.
Then, by definition, $F^\m=p-R$.
The main point is as follows:
a walk touching the wall touches it with probability one at least once at
a point with integer coordinates.
If the walk starts to the right this is clear, because $\w M=1$,
so passing through an integer point is the
only way to touch the wall coming from the right.
A walk starting to the left and touching the wall
either does it for the first time at an integer point (and we are done),
or it crosses it at some point from left to right, and then with probability
one
touches the wall again later coming from the right,
this time at an integer point,
because this is the only way to touch from the right.
In particular, $R$ is also the probability to start
to the left and touch the wall at an integer point.
Now, if we pick an integer point on the wall
and compute the probability that a walk starts to the left and passes
though this point versus the probability that a walk starts on the
right and passes though this point,
the $p$-dependent weight is the same for both, and what remains in the
ratio is the quotient of two binomial coefficients, giving $M$.
Although a walk can pass through several integer points,
as this ratio is the same for any integer point,
this is enough to show that $R$ is exactly $M$ times the probability to start
to the right and cross the wall.
The latter is $q$, so $R=Mq$ and $F^\m=p-R=p-Mq=Np-M$,
in agreement with eq.~(\ref{jd1gen}).

To finish the combinatorial analysis, we show that $Q=q$,
again in agreement with eq.~(\ref{jd1gen}).
This is a consequence of an observation made at the beginning
of sec.~\ref{sec:algtrickdual}, and summarized in eq.~(\ref{jfid}).
The probability $Q$ that a walk crosses the wall and does
it for the first time at a point with integer coordinates is the same
for right and left walks.
But for right walks $Q=q$,
because any walk starting to the right has to cross the wall,
and can only do so at an integer point.

The critical behavior of the quantities given
in eqs.~(\ref{jd1k})--(\ref{jd1gen}) is characterized by the amplitudes
\beqa
\Phi^\m_k&=&k\left(\frad{N}{N-1}\right)^{k+1},\quad
\Phi^\p_k=(k+1)\left(\frad{N}{N-1}\right)^{k+2},\nonumber\\
\Phi^\m&=&N,\quad\Phi^\p=\frad{N^2}{N-1},\quad
C^\m=\left(\frad{N-1}{2}\right)^{1/2},\quad
C^\p=\frad{N}{(2(N-1))^{1/2}}.
\label{jd1crit}
\eeqa

The integers $A^\pm_{KM+k}$ can again be evaluated
as contour integrals of the generating series $F^\pm_k(t)$.
In analogy with eq.~(\ref{j1ak}),
the expressions~(\ref{jd1k}),~(\ref{jd1fk}) yield
\beqa
A^\m_{KM+k}&=&\oint\frac{\d t}{2\pi\i t^{K+1}}\,\frac{1}{p^k}
=\oint\frac{\d p}{2\pi\i p^{KM+k+1}}\,\frac{M-Np}{(1-p)^{K+1}},\nonumber\\
A^\p_{KM+k}&=&\oint\frac{\d t}{2\pi\i t^{K+1}}\,\frac{1}{p^{k+1}}
=\oint\frac{\d p}{2\pi\i p^{KM+k+2}}\,\frac{M-Np}{(1-p)^{K+1}},
\label{jd1ak}
\eeqa
(remember that $M=N-1$), and finally
\beq
A^\m_{KM+k}=\frac{k(KN+k-1)!}{K!\,(KM+k)!},\qquad
A^\p_{KM+k}=\frac{(k+1)(KN+k)!}{K!\,(KM+k+1)!}.
\label{jdabin}
\eeq
The corresponding amplitudes $B^\pm_{KM+k}$ read asymptotically
\beq
\matrix{
B^\m_{KM+k}\approx B^\m(k/M)=k\,(M+1)^{-k/M}\hfill&(k=1,\dots,M),\hfill\cr\cr
B^\p_{KM+k}\approx B^\p(k/M)=\frad{(k+1)(M+1)^{1-k/M}}{M}
\hfill&(k=0,\dots,M-1).\hfill\cr
}
\eeq
These expressions agree with eq.~(\ref{jbper}),
since $k/M=\Frac^\pm(k/p_c)$ in both cases.
We have in particular $B^\m(1)=M/N=p_c$, and $B^\p(0)=N/M=1/p_c$.

For $M$ large, both amplitude functions $B^\pm(x)$ exhibit the scaling form
\beq
B^\pm(x)\approx x\,M^{1-x}\qquad(0\le x\le1,\quad M\gg1).
\eeq
This expression takes its maximal value,
$B^\pm\max\approx M/(\e\ln M)$, for $x\max=1/(\ln M)$.
This estimate is only by a factor $\ln M$ below
the upper bound of eq.~(\ref{bbounds}), which reads $1/q_c=N=M+1$.

\subsection{A case study: the slope $v=1/5$ or $p_c=2/5$,
i.e., $M=2$, $\w M=3$}

In this section we study the simplest case of a rational slope
which belongs to neither of the classes studied before,
namely $M=2$ and $\w M=3$.
This example corresponds to the slope $v=1/5$.
We have $N=5$, $p_c=2/5$, and $t_c=108/3125$.
We shall apply successively the three techniques
exposed in sec.~\ref{sec:algtrick}.

We start with the approach developed in sec.~\ref{sec:algcont},
based on the continuity of the path.
The functions involved in the linear system~(\ref{G=GF}) read
\beq
\matrix{
\displaystyle{G_1^\pm=\sum_{k\ge0}\bin{5k+2}{2k}t^k},\hfill&
\displaystyle{G_2^\p=\sum_{k\ge0}\bin{5k+5}{2k+1}t^k},\hfill&
\displaystyle{G_2^\m=\sum_{k\ge0}\bin{5k+4}{2k+1}t^k},\hfill\cr\cr
\displaystyle{G_{11}^\pm=G_{22}^\pm=\sum_{k\ge0}\bin{5k}{2k}t^k},\hfill&
\displaystyle{G_{12}^\p=\sum_{k\ge1}\bin{5k-3}{2k-1}t^k},\hfill&
\displaystyle{G_{12}^\m=\sum_{k\ge1}\bin{5k-2}{2k-1}t^k},\hfill\cr\cr
\displaystyle{G_{21}^\p=\sum_{k\ge0}\bin{5k+3}{2k+1}t^k},\hfill&
\displaystyle{G_{21}^\m=\sum_{k\ge0}\bin{5k+2}{2k+1}t^k}.\hfill
}
\eeq
The solution of this system leads to the generating series $F^\pm_k(t)$
of eq.~(\ref{defF}) in the form
\beq
F^\pm_1=\frac{G^\pm_1G^\pm_{22}-G^\pm_2G^\pm_{12}}
{G^\pm_{11}G^\pm_{22}-G^\pm_{21}G^\pm_{12}},\qquad
F^\pm_2=\frac{G^\pm_2G^\pm_{11}-G^\pm_1G^\pm_{21}}
{G^\pm_{11}G^\pm_{22}-G^\pm_{21}G^\pm_{12}}.
\eeq
The Taylor expansions of these quantities lead to the integers $A^\pm_k$,
namely
\beqa
F^\p_1&=&1+7\,t+99\,t^{2}+1768\,t^{3}+35530\,t^{4}+766935\,t^{5}+17368680\,
t^{6}+407139120\,t^{7}\nonumber\\
&+&9794689506\,t^{8}+240455164510\,t^{9}+5999744185435\,t^{10}+\cdots,
\nonumber\\
F^\p_2&=&2+23\,t+377\,t^{2}+7229\,t^{3}+151491\,t^{4}+3361598\,t^{5}+
77635093\,t^{6}\nonumber\\
&+&1846620581\,t^{7}\!+\!44930294909\,t^{8}\!+\!1113015378438\,t^{9}
\!+\!27976770344941\,t^{10}\!+\!\cdots,\nonumber\\
F^\m_1&=&1+5\,t+66\,t^{2}+1144\,t^{3}+22610\,t^{4}+482885\,t^{5}+10855425\,
t^{6}+253086480\,t^{7}\nonumber\\
&+&6063379218\,t^{8}+148365952570\,t^{9}+3692150267960\,t^{10}+\cdots,
\nonumber\\
F^\m_2&=&2+19\,t+293\,t^{2}+5452\,t^{3}+112227\,t^{4}+2460954\,t^{5}+
56356938\,t^{6}\nonumber\\
&+&1332055265\,t^{7}+32251721089\,t^{8}+795815587214\,t^{9}
+19939653287183\,t^{10}+\cdots\nonumber\\
\eeqa
An arbitrary number of terms can be evaluated with the help of a computer.
We have, however, found no simple closed form for the coefficients $A^\pm_k$,
such as eq.~(\ref{jabin}) for $M=1$ or eq.~(\ref{jdabin}) for $\w M=1$.

We now turn to the algebraic trick of sec.~\ref{sec:algpro}.
For $0<t<t_c$, the polynomial equation~(\ref{jupoly}), i.e., $u^2(1-u)^3=t$,
has two small solutions $u_1$ and $u_2$, such that $u_2<0<u_1<2/5$.
The relation between $t$ and any of the $F^\pm_k$
can be obtained by means of an algebraic elimination of $u_1$ and $u_2$
between eqs.~(\ref{jupoly}) and~(\ref{jlinrat}),
a tedious task that we prefer to leave to the computer.
We thus obtain
\beqa
0&=&t^6(F^\p_1)^{10}-t^4(F^\p_1)^7-11t^3(F^\p_1)^5-t^2(F^\p_1)^4-7t(F^\p_1)^2+
F^\p_1-1,\nonumber\\
0&=&t^{11}(F^\p_2)^{10}+10t^{10}(F^\p_2)^9+45t^9(F^\p_2)^8
+t^7(120t+1)(F^\p_2)^7\nonumber\\
&+&6t^6(35t+1)(F^\p_2)^6+t^5(252t+17)(F^\p_2)^5+30t^4(7t+1)(F^\p_2)^4
\nonumber\\
&+&5t^3(24t+7)(F^\p_2)^3+t^2(45t+26)(F^\p_2)^2+(10t^2+11t-1)F^\p_2+t+2,
\nonumber\\
0&=&t^4(F^\m_1)^{10}-3t^3(F^\m_1)^8+3t^2(F^\m_1)^6+11t^2(F^\m_1)^5-t(F^\m_1)^4
-4t(F^\m_1)^3+F^\m_1-1,\nonumber\\
0&=&t^9(F^\m_2)^{10}-9t^8(F^\m_2)^9+36t^7(F^\m_2)^8-84t^6(F^\m_2)^7
+126t^5(F^\m_2)^6\nonumber\\
&-&2t^4(t+63)(F^\m_2)^5+3t^3(3t+28)(F^\m_2)^4-t^2(17t+36)(F^\m_2)^3\nonumber\\
&+&t(17t+19)(F^\m_2)^2-(9t+1)F^\m_2+t+2.
\label{jalgft}
\eeqa

The algebraic relations between $F^\pm$ and $p$ are slightly more complicated.
We assume that $p>2/5$.
The elimination of $t$ and $u_1$ and $u_2$ between
eqs.~(\ref{jupoly}),~(\ref{jlinrat}), and~(\ref{clauderat}) leads to
\beqa
0&=&p^3(F^\p)^6-10p^3(F^\p)^5+3p^2(15p-1)(F^\p)^4-p(5p-1)(25p-1)(F^\p)^3
\nonumber\\
&+&3p(15p-1)(5p-2)(F^\p)^2-10p(5p-2)^2F^\p+(5p-2)^3,\nonumber\\
0&=&(F^\m)^6-(10p-9)(F^\m)^5+3(p-1)(15p-11)(F^\m)^4\nonumber\\
&-&(p-1)(125p^2-185p+63)(F^\m)^3+3(p-1)^2(15p-11)(5p-2)(F^\m)^2\nonumber\\
&-&(p-1)^2(10p-9)(5p-2)^2F^\m+(p-1)^3(5p-2)^3.
\label{jalgfp}
\eeqa

The above algebraic relations~(\ref{jalgft}),~(\ref{jalgfp})
have respective degrees $\deg_t(2,5)=10$ in the $F^\pm_k$
and $\deg_p(2,5)=6$ in $F^\pm$, in agreement with eq.~(\ref{jdegre}).
These relations also allow an investigation of the critical behavior
of quantities as $p\to 2/5$.
It turns out that all the critical amplitudes defined in sec.~\ref{sec:algcrit}
can be expressed as rational functions of one single irrational number:
\beq
\z=10^{1/3},
\eeq
namely
\beqa
(F^\p_1)_c&=&\frad{125}{108}(2+2\z-\z^2)=1.929723,\qquad
(F^\p_2)_c=\frad{3125}{1944}(2-10\z+5\z^2)=5.889270,\nonumber\\
(F^\m_1)_c&=&\frad{25}{18}(\z-1)=1.603382,\qquad
(F^\m_2)_c=\frad{625}{108}(3-\z)=4.893318,\nonumber\\
\Phi^\p_1&=&\frad{625}{108}=5.787037,\qquad
\Phi^\p_2=\frad{78125}{11664}(4-2\z+\z^2)=29.020380,\nonumber\\
\Phi^\m_1&=&\frad{125}{36}=3.472222,\qquad
\Phi^\m_2=\frad{3125}{648}(2+\z)=20.034890,\nonumber\\
\Phi^\p&=&\frad{5}{18}(10-2\z+\z^2)=2.870200,\qquad
\Phi^\m=\frad{1}{3}(5+\z)=2.384812,\nonumber\\
C^\p&=&\frad{1}{6\sqrt3}(10-2\z+\z^2)=0.994266,\qquad
C^\m=\frad{1}{5\sqrt3}(5+\z)=0.826123.
\eeqa

To close up the study of this example,
let us consider the duality approach of sec.~\ref{sec:algtrickdual}.
We have $I_0^+=\{1,3\}$ and $I_0^-=\{1,2,4\}$.
We again focus our attention on quantities with the superscript $\p$.
Eqs.~(\ref{dualF1}), (\ref{dualFl}) yield four equations for the four unknown
functions $\oF^\p_1,\dots,\oF^\p_4$:
\beqa
\oF^\p_1&=&1+t\left(\oF^\p_1\oF^\p_4+\oF^\p_2\oF^\p_3\right),\nonumber\\
\oF^\p_2&=&{(\oF^\p_1)}^2+t\oF^\p_3\oF^\p_4,\nonumber\\
\oF^\p_3&=&\oF^\p_1\oF^\p_2,\nonumber\\
\oF^\p_4&=&\oF^\p_1\oF^\p_3.
\eeqa
This non-linear system can be solved in closed form.
The trick is to define a variable $x=\oF^\p_2/{(\oF^\p_1)}^2$.
The functions $\oF^\p_k$ and the variable $t$ can then be expressed
as rational functions of $x$:
\beqa
t&=&\frad{x^3(x-1)}{(x^2+x-1)^5},\nonumber\\
\oF^\p_1&=&\frad{x^2+x-1}{x},\nonumber\\
\oF^\p_2&=&\frad{(x^2+x-1)^2}{x},\nonumber\\
\oF^\p_3&=&\frad{(x^2+x-1)^3}{x^2},\nonumber\\
\oF^\p_4&=&\frad{(x^2+x-1)^4}{x^3}.
\eeqa

In this case, the Riemann surface of $t$ and of the
$\oF^\p_1,\dots,\oF^\p_4$ has genus 0.
It is a branched covering of the $t$-sphere of degree $\deg_t(2,5)=10$,
with $x$ being the uniformizing variable.
The physical region $0\le t\le t_c=108/3125$ corresponds
to $1\le x\le x_c=(2+\z^2)/6=1.106931$, with $\d t/\d x=0$ at $x=x_c$.
We close up this section by observing that in general
the genus of the Riemann surface where all the functions $\oF^\p_k$ and $t$
are uniform grows very rapidly with $M$ and $\w M$.

\subsection{Directed scaling limit: $v\to1$, i.e., $p_c\to0$, and $p\to0$}

We now investigate the situation where $p$ and $p_c$ are simultaneously small.
We refer to this case as the `directed scaling limit',
since the random walks are almost perfectly directed toward the rightmost limit
$(x=t)$ of phase space~[see Figure~1].
This non-conventional scaling limit turns out to be characterized
by non-trivial scaling laws, in the three regimes described in section~2.

We introduce the scaling variable
\beq
\xi=\frac{p}{p_c}.
\eeq
We consider first the convergent regime $(\xi>1)$.
The explicit results derived in sec.~\ref{sec:vmun}
for the family of rational slopes with $M=1$
exhibit scaling behavior as $p_c\ll1$, i.e., $N\gg1$.
Indeed the small root $u_1$ of eq.~(\ref{jupoly}) scales as
$u_1\approx\xi_1p_c$, where $0<\xi_1<1$ is related to $\xi>1$ by
\beq
\xi_1\e^{-\xi_1}=\xi\e^{-\xi}.
\label{jexpconj}
\eeq
Eqs.~(\ref{j1f}),~(\ref{j1sol}) then yield
\beqa
&&\!\!\!F_1^\pm\approx\e^{\xi_1},\\
&&\!\!\!F^\pm\approx(\xi-\xi_1)p_c,
\label{jscaf}\\
&&\Pi\approx\xi_1(\xi-\xi_1)p_c^2.
\eeqa

More generally, the algebraic formalism of sec.~\ref{sec:algpro}
can be worked out for an arbitrary rational slope $p_c=M/N$
in the directed scaling limit, i.e., $M$ finite and $N\gg1$.
The $M$ small roots $u_\alpha$ of eq.~(\ref{jupoly}),
with $\alpha=1,\dots,M$, now scale as $u_\alpha\approx\xi_\alpha p_c$,
where $\xi_\alpha$ is the solution such that $\abs{\xi_\alpha}<1$ of
\beq
\xi_\alpha\e^{-\xi_\alpha}=\o^{\alpha-1}\xi\e^{-\xi},
\eeq
with the definition~(\ref{jdefo}).
As a consequence, solving the linear system~(\ref{jlinrat})
amounts to inverting the discrete Fourier-transform matrix
$S_{k,\ell}=\o^{k\ell}$ with $k,\ell=1,\dots,M$.
We have $(S^{-1})_{k,\ell}=\o^{-k\ell}/M$, so that finally
\beq
F^\pm_k\approx p_c^{1-k}(\xi\e^{-\xi})^{-k}
\,\frac{1}{M}\sum_{\alpha=1}^M\o^{-k(\alpha-1)}\xi_\alpha.
\eeq

The scaling form~(\ref{jscaf}) for the survival probabilities $F^\pm$
thus holds true for any (rational) slope in the directed scaling limit.
The probability $\Pi$ still scales as $p_c^2$,
but with a non-trivial coefficient:
\beq
\Pi\approx(\xi-\xi_1)\left(\frac{1}{M}\sum_{\alpha=1}^M\xi_\alpha\right)p_c^2.
\eeq

The critical behavior of the survival probabilities
as $p\to p_c$, i.e., $\xi\to1$, is characterized by the amplitudes
\beq
\Phi^\pm\approx 2.
\label{jcasy}
\eeq
We thus obtain the scaling behavior of the amplitudes $C^\pm$
of eq.~(\ref{jfc}) in the marginal regime $(p=p_c\ll1$ or $\xi=1)$:
\beq
C^\pm\approx(2p_c)^{1/2},
\eeq
in agreement with eq.~(\ref{jylim}).

Some of the above results can be alternatively derived by means
of the integers $A^\pm_k$.
The bounds~(\ref{bounds}) imply that these numbers exhibit a simple
scaling as $p_c\ll1$, namely
\beq
A^\pm_k\approx\frac{(k/p_c)^{k-1}}{k!}.
\label{jcla}
\eeq
The behavior of the $A^\pm_k$ for large $k$ agrees with eq.~(\ref{jasya}),
with a trivial modulation $B^\pm_k\approx1$.

Eq.~(\ref{claude}) then yields
\beq
F^\pm\approx\left(\xi-\phi\left(\xi\e^{-\xi}\right)\right)p_c,
\label{jfh}
\eeq
with
\beq
\phi(z)=\sum_{k\ge1}\frac{k^{k-1}z^k}{k!}.
\eeq
This functional relation between $\phi$ and $z$ is equivalent to
\beq
z=\phi\e^{-\phi},
\label{jhansen}
\eeq
with the condition that $\phi\to0$ for $z\to0$.
The identity (\ref{jhansen}), which is given in refs.~\cite{GKP,H},
can be checked by means of the contour integrals
\beq
\oint\frac{\d z}{2\pi\i z^{k+1}}\,\phi(z)
=\oint\frac{\d\phi}{2\pi\i\phi^k}\,(1-\phi)\e^{k\phi}
=\frac{k^{k-1}}{k!}.
\eeq
In eq.~(\ref{jfh}) we have $z=\xi\e^{-\xi}$ with $\xi>1$,
so that eq.~(\ref{jexpconj}) leads to the identification $\phi=\xi_1$.
The result~(\ref{jscaf}) is thus recovered,
without recourse to the algebraic trick.
Its validity for any slope in the directed scaling limit,
either rational or not, is thus established.

\vskip 8.5cm{\hskip 0.8cm}
\includegraphics{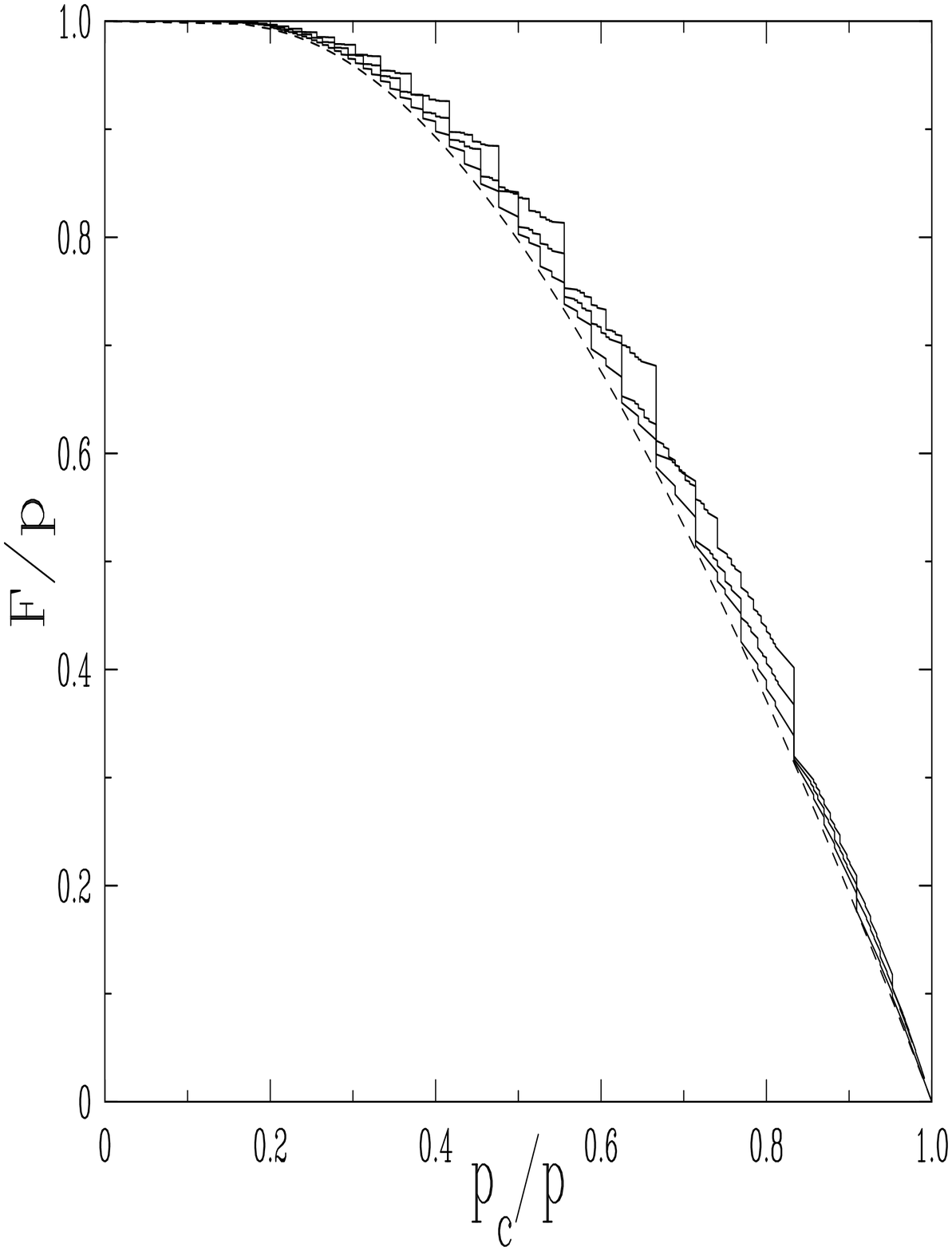}

\capt{16}{J3}{Plot of $F/p$ against $p_c/p$ for $p=0.1$, 0.2, and 0.3,
illustrating the scaling law~(\ref{jcls})
pertaining to the directed scaling limit, shown as a dashed curve.}

The result~(\ref{jscaf}) can be recast
as the following scaling relation for the survival probabilities
\beq
\frac{F^\pm}{p}\approx1-\exp\left(-\frac{F^\pm}{p_c}\right),
\label{jcls}
\eeq
which is valid throughout the convergent regime of the directed scaling limit,
i.e., when $p$, $p_c<p$, and $F^\pm$ are simultaneously small.
Figure~16 illustrates this relation.
The ratio $F/p$ is plotted against $p_c/p$.
The data for $p=0.3$, 0.2, and 0.1 (full curves, from top to bottom),
already presented in Figure~10,
are found to smoothly converge toward the scaling law~(\ref{jcls}),
shown as a dashed curve.

To close up this section,
we consider briefly the large-deviation regime $(p<p_c$ or $\xi<1)$.
When $p$ and $p_c$ are simultaneously small with $p<p_c$,
the survival probabilities $F^\pm(n,v)$ can be estimated
by inserting the expression~(\ref{jcla}) for the integers $A^\pm_k$
into eq.~(\ref{jcn2}).
We obtain after some algebra an asymptotic expression
of the form~(\ref{jasyf}),
where the entropy function scales as $S(v)\approx p_c(\xi-1-\ln\xi)$,
as it should, and where the periodic amplitudes $a^\pm(x)$ and $b^\pm(x)$ read
\beq
a^\pm(x)\approx\frac{\xi^{1-\Frac^\mp(x)}}{(2\pi p_c)^{1/2}(1-\xi)},\qquad
b^\pm(x)\approx\frac{\xi^{1-\Frac^\mp(x)}}{(2\pi p_c)^{1/2}
\left(1-\xi\e^{1-\xi}\right)}.
\eeq

\section{Discussion}
\label{sec:dis}

We have performed a detailed analysis of the statistics
of persistent events in the case of the one-dimensional
lattice random walk
in the presence of an obstacle moving ballistically with velocity $v$.
Both space and time are discrete,
so that the underlying lattice structure yields
a highly non-trivial dependence on the velocity $v$ of the obstacle,
with discontinuities at rational values of $v$,
for most of the quantities investigated in this work.
This is especially the case for
the limit survival probabilities $F^\pm$ in the convergent regime,
and for the amplitudes $C^\pm$ characteristic of the marginal regime,
respectively shown in Figures~10 and 14--15.
We have obtained a deep insight into the problem by means of the algebraic
methods exposed in sec.~5.

We want to emphasize that the statistics of persistent events
can be investigated in a broader perspective,
even within the realm of one-dimensional random walks,
constrained to remain on the left of a moving obstacle.
Consider a general random walk consisting of independent identically
distributed steps $\eps_n$,
and a moving obstacle, whose position is given by an arbitrary function $X(t)$.
The following situations illustrate the richness of possible behavior
of this seemingly simple system [see refs.~\cite{DG,BBDG,B,KR} for related
questions].

\begin{itemize}

\item{
The steps $\eps_n$ have a distribution whose first two moments are finite,
and the obstacle moves ballistically: $X(t)=vt$.
The Sparre Andersen formalism of sec.~\ref{sec:sa} applies
in this situation.
The survival probability is described by the scaling
laws~(\ref{jasyf}), (\ref{jfc}), and~(\ref{jfcv}),
associated with the three regimes described in the introduction.
The entropy function $S$, the constant $C$ of the marginal regime,
and the limit survival probability $F$ in the convergent regime,
are non-universal,
in the sense that they depend on details of the distribution of the steps.
The standard large-deviation formalism allows to determine $S$
by means of its Legendre transform.
The constant $C$ and the limit survival probability $F$
are given by the formal expressions~(\ref{jc}) and~(\ref{jf}).
No closed-form expressions for these quantities are known in general.
Let us mention, however,
that the second expression of eq.~(\ref{jfpc}) is quite general,
with ${\rm Var}\ \eps=\mean{\eps^2}-\mean{\eps}^2$ replacing $1-v^2$.
}

\item{
The steps $\eps_n$ have a symmetric distribution whose second moment is finite,
and the obstacle moves according to $X(t)=g\,t^{1/2}$,
in units where $\mean{\eps^2}=1$.
In this case the survival probability of the walker in the presence of
the obstacle exhibits a scaling behavior of the form
\beq
F(n,g)\sim n^{-\w\theta(g)}\qquad(n\gg1),
\label{jthdif}
\eeq
where the exponent $\w\theta(g)$ is a non-trivial universal,
continuously decreasing function of $g$,
given in terms of a zero of the parabolic cylinder function~\cite{B,KR}.
The marginal regime is recovered as $\w\theta(0)=1/2$.
The result~(\ref{jthdif}) crosses over to the large-deviation regime
as $g\to-\infty$,
where the exponent diverges as $\w\theta(g)\approx g^2/8$,
while it crosses over to the convergent regime as $g\to+\infty$,
where the exponent vanishes as
$\w\theta(g)\approx g\,\e^{-g^2/2}/(8\pi)^{1/2}$.
The exponent $\w\theta(g)$ is related to the persistence exponent
$\theta(g)$ defined in the introduction, for the sake of consistency
with refs.~\cite{DG,BBDG,Drouffe}, by $\w\theta(-g)=\theta(g)$.
}

\item{
The steps $\eps_n$ have a symmetric broad (L\'evy) distribution,
with long tails falling off as $\rho(\eps)\sim\abs{\eps}^{-\mu-1}$,
with $0<\mu<2$,
and the position of the obstacle obeys an asymptotic long-time behavior
of the form $X(t)\approx g\,t^{1/\mu}$.
Then the survival probability is expected to exhibit a scaling behavior
of the form~(\ref{jthdif}),
again with a continuously varying exponent $\w\theta(g)$,
that is universal if the $\eps_n$ are measured in appropriate units,
but whose expression is not known in general.

A simple example in this category
corresponds to steps having a Cauchy distribution:
$\rho(\eps)=1/\left(\pi(1+\eps^2)\right)$,
with an obstacle moving ballistically, according to $X(t)=vt$.
The Sparre Andersen formalism again applies to this situation.
The stability of the Cauchy law under convolution implies that
the one-time distribution function reads
\beq
P(n,v)=P(v)=\int_{-\infty}^{v}\frac{\d\eps}{\pi(1+\eps^2)}
=\frac{1}{2}+\frac{1}{\pi}\arctan v,
\eeq
independently of $n$.
Eq.~(\ref{jic})
then leads to $f(z,v)=(1-z)^{-P(v)}$, hence
\beq
F(n,v)=\frac{\Gamma\bigl(n+P(v)\bigr)}{n!\,\Gamma(P(v))}
\approx\frac{n^{-\w\theta(v)}}{\Gamma(P(v))}\qquad(n\gg1),
\eeq
where $\w\theta(v)$ is again a universal exponent, given by
\beq
\w\theta(v)=1-P(v)=\frac{1}{2}-\frac{1}{\pi}\arctan v.
\label{jthcau}
\eeq
The marginal regime is again recovered as $\w\theta(0)=1/2$.
The result~(\ref{jthcau}) crosses over to the convergent regime
as $v\to+\infty$, where the exponent vanishes as
$\w\theta(v)\approx1/(\pi v)$.
Contrary to the previous situation,
the exponent does not cross over to the large-deviation regime
as $v\to-\infty$, since it admits a finite limit $\w\theta(-\infty)=1$.
The exponent $\w\theta(v)$ is again related to the persistence exponent
$\theta(v)$ defined in the introduction by $\w\theta(-v)=\theta(v)$,
yielding in the present case $\theta(v)=P(v)$.
}

\end{itemize}

\subsubsection*{Acknowledgments}

It is a pleasure for us to thank Vincent Lafforgue
for his constant interest in our work,
and for his illuminating ideas on two points of this problem,
reproduced in secs.~\ref{sec:algpro} and~\ref{sec:vwemun}.
We also warmly acknowledge interesting correspondence with Don Zagier.

\newpage
\appendix
\section{Combinatorial proof of the Sparre Andersen identity~(\ref{jic})}

In this Appendix we provide an elementary and self-contained
combinatorial proof of the Sparre Andersen identity~(\ref{jic}).

We keep notations consistent with the body of the paper,
dropping for the sake of simplicity the $\pm$ superscript
and the dependence on $v$.
Let $\{\eps_n\}_{n=1,2,\dots}$ be independent identically distributed
random variables, and let $x_0=0$, $x_1=\eps_1$, $\dots$,
$x_n=\eps_1+\cdots+\eps_n$, $\dots$ be their partial sums.
For $n=1,2,\dots$ we denote by $P(n)=\prob\big\{x_n \ge 0\big\}$
the probability that the $n$-th partial sum is non-negative,
and by $F(n)=\prob\big\{x_1 \ge 0,\dots,x_n \ge 0\big\}$
the probability that the first $n$ partial sums are non-negative.
We set consistently $F(0)=1$.

We want to prove that the generating functions
\beq
p(z)=\sum_{n\ge1}\frac{P(n)}{n}z^n,\qquad f(z)=\sum_{n\ge0}F(n)z^n
\eeq
are related by the Sparre Andersen identity~(\ref{jic}), i.e.,
\beq
f(z)=\exp(p(z)).
\label{appsa}
\eeq
For each $n\ge 1$,
we define the event ${\bf A}_n$ and $n$ events ${\bf B}_n^i$
for $i=1,\dots,n$ by
\beq
\begin{array}{rcl}
{\bf A}_n&=&\{\eps_1+\cdots+\eps_n\ge 0\},\nonumber\\
{\bf B}_n^i&=&\{\eps_i \ge 0, \eps_i+\eps_{i+1} \ge
0,\dots,\eps_i+\cdots+\eps_n \ge 0,\nonumber\\
& &{\hskip 2mm}\eps_i+\cdots+\eps_n+\eps_1 \ge
0, \dots, \eps_i+\cdots+\eps_n+\eps_1+\cdots+\eps_{i-1} \ge 0\}.
\end{array}
\eeq

The first observation is that ${\bf A}_n=\bigcup_{i=1}^{n} {\bf B}_n^i$.
It is clear from the definition that ${\bf B}_n^i \subset{\bf A}_n$.
To prove the reverse inclusion, assume ${\bf A}_n$ holds, i.e., $x_n\ge 0$.
Let $i \in \{1,\dots,n\}$ be such that $x_{i-1}$ is the minimum of
$\{x_0,\dots,x_{n-1}\}$
(in case of degeneracy, we take the smallest such $i$).
We claim that ${\bf B}_n^{i}$ is realized.
Indeed, by definition of $m$,
$x_m-x_{i-1} \ge 0$ for $m \in \{i,\dots,n-1\}$,
and (as $x_n \ge 0$), $x_n+(x_m-x_{i-1}) \ge 0$ for $m \in
\{0,\dots,i-1\}$: this covers the definition of ${\bf B}_n^{i}$.

We apply the inclusion-exclusion principle and write
\beq
P(n)=\prob\big\{{\bf A}_n\big\}=\sum_{k=1}^n(-)^{k+1}\sum_{0<i_1<\dots
<i_k<n+1}\prob\big\{{\bf B}_n^{i_1}\cap\dots\cap{\bf B}_n^{i_k}\big\}.
\label{appp}
\eeq

Now comes the second observation.
Choose $k\ge1$ and $0<i_1<\dots<i_k<n+1$.
Then
\beqa
{\bf B}_n^{i_1}\cap\dots\cap{\bf B}_n^{i_k}
&=&\{\eps_{i_1}\ge0,\dots,\eps_{i_1}+\cdots+\eps_{i_2-1}\ge0\}\nonumber\\
&&\cap\dots\cap\{\eps_{i_j}\ge0,\dots,\eps_{i_j}+\cdots+\eps_{i_{j+1}-1}\ge0\}
\nonumber\\
&&\cap\dots\cap\{\eps_{i_k}\ge0,\dots,\eps_{i_k}+\cdots+\eps_{n}\ge0,\dots,
\nonumber\\
&&{\hskip 15.3mm}\eps_{i_k}+\cdots+\eps_{n}+\eps_1+\cdots+\eps_{i_1-1}\ge0\}.
\eeqa
Indeed all the inequalities on the right-hand-side are among the defining
inequalities for the left-hand-side.
Moreover, it is obvious that the
missing inequalities on the right-hand-side are sums of the written ones.
We have thus succeeded in decomposing ${\bf B}_n^{i_1} \cap \dots
\cap {\bf B}_n^{i_k}$ as a product of $k$ events.
These events are independent because they
involve distinct steps of the random walk.
Another nice feature is that the event
$\{\eps_{i_j}\ge 0, \dots, \eps_{i_j}+\cdots+\eps_{i_{j+1}-1}\ge 0\}$
has probability $F(i_{j+1}-i_{j})$.
Indeed, the steps are identically distributed,
so that exchanging the steps do not change probabilities, and
$\{\eps_{i_j}\ge 0, \dots,\eps_{i_j}+\cdots+\eps_{i_{j+1}-1}\ge 0\}$
has the same probability as
$\{\eps_1\ge 0,\dots,\eps_1+\cdots+\eps_{i_{j+1}-i_j}\ge 0\}$.
We can thus rewrite eq.~(\ref{appp}) for $P(n)$ as
\beq
P(n) = \sum_{k=1}^n (-)^{k+1} \sum_{0 < i_1 < \dots
< i_k < n+1} F(i_2-i_1) \dots F(i_k-i_{k-1})F(i_1-i_k+n).
\label{apppq}
\eeq

The last step of the proof is as follows.
Multiply eq.~(\ref{apppq}) by $z^n$ and sum over $n=1,2,\dots$
The left-hand-side yields $zp'(z)$,
where the accent denotes a differentiation.
At fixed $k$, the $k$-uple sum in the right-hand-side can be performed
by taking $j_1=i_2-i_1\ge1$, $j_2=i_3-i_2\ge1$, $\dots$,
$j_{k-1}=i_k-i_{k-1}\ge1$,
$j_k=i_1-i_k+n\ge1$ as independent summation indices.
The power of $z$ can be recast as $z^{j_1+\cdots+j_k}$.
Each of the sums over $j_1$, $\dots$, $j_{k-1}$ brings a factor $f(z)-1$,
while the last sum over $j_k$ yields $zf'(z)$.
We are thus left with the equation
\beq
zp'(z)=\sum_{k\ge1}(1-f(z))^{k-1}\,zf'(z),
\eeq
i.e.,
\beq
p'(z)=\frac{f'(z)}{f(z)},
\eeq
which yields the identity~(\ref{appsa}) by integration,
since $p(0)=0$, $f(0)=1$.

\newpage
\section*{Figure captions}

\capt{1}{M1}{Configuration space of the problem.}

\capt{2}{M2}{A walk remaining on the left of the wall,
thus contributing to the survival probability.}

\capt{3}{M3}{A walk crossing the wall,
thus not contributing to the survival probability.}

\capt{4}{J4}{Periodic amplitudes $a(x)$ and $b(x)$,
for $p=0.3$ and $v$ equal to the golden slope.}

\capt{5}{M4}{Labeling of crossing edges by an integer $k=1$, 2, 3,
equal to the number of steps of the walker to the left.}

\capt{6}{M5}{Combinatorial approach: truncated Pascal triangle.}

\capt{7}{M6}{The probability flow is characterized by integers $A^\pm_k$,
living on the crossing edges [cf. Figure~5].
In the example, we have $A^\pm_1=1$, $A^\pm_2=2$, $A^\pm_3=5$, and so on.}

\capt{8}{M7}{The canonical walk.}

\capt{9}{J1}{Roots of the polynomial equation~(\ref{jupoly})
for $M=4$, $N=11$, and $p=1/2$.}

\capt{10}{J2}{Plot of the survival probabilities $F^\pm$ against $v$,
for several values of $p$, indicated on the curves.
The apparent $v\to1$ limit of $F(v)$ is $F^-(1)=p$~[see eq.~(\ref{jsimple})],
whereas the jump at $v=1$ to $F^+(1)=1$ is not visible.}

\capt{11}{M9}{Combinatorial proof of the identity~(\ref{jfid}).}

\capt{12}{J7}{Plot of the periodic amplitude $B(x)$,
for the golden slope.}

\capt{13}{J8}{Plot of the periodic amplitude $\beta(\theta)$,
for the golden slope.}

\capt{14}{J5}{Logarithmic plot of the critical amplitude $C(v)$ against $v$.}

\capt{15}{J6}{Plot of $y(v)$, defined in eq.~(\ref{jy}), against $v$.}

\capt{16}{J3}{Plot of $F/p$ against $p_c/p$ for $p=0.1$, 0.2, and 0.3,
illustrating the scaling law~(\ref{jcls})
pertaining to the directed scaling limit, shown as a dashed curve.}
\end{document}